\documentclass[twocolumn]{aastex63}
\usepackage{longtable}

\makeatletter

\newcommand{\Rmnum}[1]{\expandafter\@slowromancap\romannumeral #1@}
\makeatother

\received{April 26, 2022}
\revised{July 21, 2022}
\accepted{}
\submitjournal{ApJ}

\shorttitle{Identification and parameter determination of F-type Herbig stars}
\shortauthors{Yun-Jin Zhang et al.}
\begin{document}

\title{Identification and parameter determination of F-type Herbig stars from LAMOST DR8}

\correspondingauthor{A-Li Luo}
\email{* lal@nao.cas.cn}

\author[0000-0001-7210-0666]{Yun-Jin Zhang}
\affiliation{CAS Key Laboratory of Optical Astronomy, National Astronomical Observatories, Beijing 100101, China}
\affiliation{University of Chinese Academy of Sciences, Beijing 100049, China}

\author[0000-0001-7865-2648]{A-Li Luo}
\affiliation{CAS Key Laboratory of Optical Astronomy, National Astronomical Observatories, Beijing 100101, China}
\affiliation{University of Chinese Academy of Sciences, Beijing 100049, China}

\author[0000-0003-3168-2617]{Biwei Jiang}
\affiliation{Department of Astronomy, Beijing Normal University, Beijing 100875, China}

\author[0000-0003-0716-1029]{Wen Hou}
\affiliation{CAS Key Laboratory of Optical Astronomy, National Astronomical Observatories, Beijing 100101, China}

\author[0000-0002-9081-8951]{Fang Zuo}
\affiliation{CAS Key Laboratory of Optical Astronomy, National Astronomical Observatories, Beijing 100101, China}

\author[0000-0003-0716-1029]{Bing Du}
\affiliation{CAS Key Laboratory of Optical Astronomy, National Astronomical Observatories, Beijing 100101, China}

\author[0000-0002-8913-3605]{Shuo Li}
\affiliation{CAS Key Laboratory of Optical Astronomy, National Astronomical Observatories, Beijing 100101, China}
\affiliation{University of Chinese Academy of Sciences, Beijing 100049, China}

\author{Yong-Heng Zhao}
\affiliation{CAS Key Laboratory of Optical Astronomy, National Astronomical Observatories, Beijing 100101, China}
\affiliation{University of Chinese Academy of Sciences, Beijing 100049, China}



\begin{abstract}
We identify 20 F-type Herbig stars and provide a list of 22 pre-main-sequence candidates from LAMOST DR8. The effective temperature, distance, extinction, stellar luminosity, mass, and radius are derived for each Herbig star based on optical spectra, photometry, Gaia EDR3 parallaxes, and pre-main-sequence evolutionary tracks. According to spectral energy distributions, 19 F-type Herbig stars belong to Class \Rmnum {2} YSOs, and one belongs to the flat-spectrum class. Four have Spitzer IRS spectra, of which three show extremely weak polycyclic aromatic hydrocarbons emissions, and three with both amorphous and crystalline silicate emissions share the similar parameters and are at the same evolutionary stage. We detect a solar-nearby outbursting EXor Herbig star J034344.48+314309.3, possible precursor of a Herbig Ae star. Intense emission lines of H\Rmnum{1}, He\Rmnum{1}, O\Rmnum{1}, Na\Rmnum{1}, and Ca\Rmnum{2} originated from the rapid accretion during the outbursts are detected in its optical spectra, and silicate emission features are detected in its infrared spectrum. We also make a statistic analysis on the disk properties of all known Herbig stars using the defined infrared spectral indices ($\alpha_{J-K_S}$ and $\alpha_{K_S-W3}$). The proportion of Herbig stars with moderate infrared excesses decreases as effective temperature increases. The majority of the precursors (F-, G-, or K- type) have moderate infrared excesses.  Hotter Herbig stars tend to have a larger proportion with large infrared excesses. The trends may be due to the fact that hotter stars have larger areas of re-emitting dust, although there is some scatter due to the particularities of each disk.
\end{abstract}



\section{Introduction} \label{sec:intro}

Young stellar object (YSO) denotes a young star in its early stage of evolution. Pre-main-sequence (PMS) stars and protostars are the two subtypes of YSOs. Protostars with collapsing envelopes are deeply embedded in the parent molecular clouds and optically invisible \citep{SED.1987IAUS..115....1L}. Compared with the younger counterparts, after dust envelopes infalling, PMS stars with accretion disks are optically visible. When approaching the main-sequence stage, the circumstellar dust disk will eventually be dispersed by planet formation and radiation pressure from the central star.

Herbig Ae/Be stars (HAeBes) and T Tauri stars (TTs) are the two main types of PMS stars. The concept of TTs was defined as PMS stars with late spectral types (later than late F, \citet{1945ApJ...102..168J,1962AdA&A...1...47H,1983A&A...126..438B}). As the counterpart of TTs, HAeBes were first defined by \citet{Herbig.1960ApJS....4..337H} as emission-line stars with a spectral type of A or earlier which are located in obscured regions and illuminate fairly bright nebulosity in the immediate vicinity. Subsequently, \citet{Herbig.1972ApJ...173..353S} extended the HAeBes definition with additional properties, including infrared (IR) excess, time variation, linear polarization, and association with star-forming regions. Since then, the HAeBes data set has been enlarged by numerous studies over decades \citep{1984A&AS...55..109F,7.1994A&AS..104..315T,HAeBe4.2003AJ....126.2971V,JHK.2005AJ....129..856H,HAeBe2.2006Ap&SS.305...11Z,HAeBe10.2006ApJ...653..657M,HAeBe9.2006MNRAS.367..737B,HAeBe8.2010AJ....139...27S,HAeBe7.2010A&A...517A..67C,HAeBe6.2013MNRAS.429.1001A,HAeBe5.2015MNRAS.453..976F,HAeBe1.2016NewA...44....1C,252HAeBe.2018A&A...620A.128V,HAeBe.2019ApJ...878..147K,2020MNRAS.494.3512M,2020A&A...638A..21V,2021AJ....162...71P,2022ApJS..259...38Z,2022ApJ...930...39V}.

In early research, HAeBes are usually limited to PMS stars with A or B spectral types. But in more recent works, HAeBes are generally defined as PMS stars with intermediate mass (1.5 or 2$-$10 $M_\odot$), and a few F-type Herbig stars at the early evolutionary stage in the PMS track have also been included. \citet{HAeBe.2019ApJ...878..147K} identified 13 HAeBes with spectral types ranging from early G to B in the Small Magellanic Cloud. \citet{2022ApJ...930...39V} identified 128 Herbig stars based on spectra and Gaia EDR3 data, the spectra of which also range from early G to B.

As the bridge of typical HAeBes and TTs, research about F-type PMS stars is rare. \citet{2004AJ....128.1294C,2017A&A...608A..77L} defined a specified intermediate-mass T Tauri (IMTT) class with 1$-$4 $M_\odot$. \citet{2021AJ....162..153N} characterized the X-Ray emission of intermediate-mass PMS stars in the Carina Nebula. Actually, IMTTs are usually the precursors of the HAeBes, often of F type. Depending on the PMS evolutionary tracks \citep{2011A&A...533A.109T,2012MNRAS.427..127B,2017ApJ...835...77M}, most of the F-type PMS stars will be able to become HAeBes eventually. To distinguish from traditional Herbig Ae/Be stars, we called these F-type PMS stars with stellar masses larger than 1.5 $M_\odot$ as F-type Herbig stars. F-type PMS with stellar masses less than 1.5 $M_\odot$ as F-type T Tauri stars. Despite the similar spectral types, ages are quite different. F-type Herbig stars are the precursors of HAeBes. \citet{2021A&A...652A.133V} pointed out the timescales on evolving toward the main sequence are comparable to those of typical disk dissipation, and a full picture of disk dispersal must include these precursors. They analysed 49 precursors of young HAeBes with spectral types ranging from F to K3 in literature, and found the polycyclic aromatic hydrocarbons (PAH) detection frequency of the precursors is different from that of HAeBes and TTs. They also suggested gaps and spirals are present around these younger precursors of HAeBes.

According to the SIMBAD (keyword TT* and Ae*), the number of known F-type PMS stars is 60, far less than that of HAeBes and TTs. From the 2.53 million F-type spectra of the Large Sky Area Multi-Object Fiber Spectroscopic Telescope (LAMOST) DR8, we have a chance to discover more F-type PMS stars. Fortunately, 20 F-type Herbig stars were identified, along with 22 PMS candidates. The effective temperature, distance, extinction, stellar luminosity, mass, and radius were derived for each Herbig stars based on optical spectra, photometry, Gaia EDR3 parallaxes, and pre-main-sequence evolutionary tracks. According to spectral energy distributions (SEDs), 19 F-type Herbig stars belong to Class \Rmnum {2} YSOs, and one belongs to the flat-spectrum class.

The outline of this paper is as follows. In Section \ref{sec:observations}, we introduce our data and our catalogs. In Section \ref{sec:selection}, we make the sample selection and confirmation. In Section \ref{sec:parameters}, we determine stellar parameters for our F-type Herbig stars. In Section \ref{sec:analysis}, we make a further analysis on the Herbig stars. In Section \ref{sec:discussion}, we detect a solar-nearby outbursting EXor Herbig star and discuss the disk properties of all known Herbig stars. A brief summary is provided in Section \ref{sec:summary}.

\section{Observations} \label{sec:observations}

LAMOST is a reflecting Schmidt telescope located at the Xinglong Station of the National Astronomical Observatory, China (40$^\circ$N,105$^\circ$E), with a mean aperture of 4.3 m and a field of view of 5$^\circ$. The LAMOST spectrograph has two resolving modes, the low-resolution mode of R$\sim$1800 and the medium-resolution mode of R$\sim$7500.
Until 2021 June, DR8\footnote{\url{http://www.lamost.org/dr8/}} published more than 11 million low-resolution spectra covering 3690$-$9100$\rm\AA$ and about 17.6 million single exposure (4.7 million coadded) medium-resolution spectra covering 4950$-$5350$\rm\AA$ and 6300$-$6800$\rm\AA$.

The data we used includes 2,538,636 low-resolution spectra of F-type stars classified by LAMOST 1D-pipeline \citep{LAMOST.2015RAA....15.1095L} based on LAMOST DR8 v1.0. All spectra have been reduced through LAMOST 2D-pipeline with telluric absorption corrected, sky background subtracted, and flux and wavelength calibrated \citep{sky.2017RAA....17...91B}.

All the catalogs can be downloaded \textbf{via China-VO: \dataset[https://nadc.china-vo.org/res/r101135/]{https://nadc.china-vo.org/res/r101135/}.}

\section{Sample Selection} \label{sec:selection}

\subsection{Criterion} \label{subsec:criterion}

The determination of circumstellar dust disk is essential for the identification of PMS stars. The main features of circumstellar disk on optical and infrared bands are the wide H$\alpha$ emission (gaseous disk) and the apparent infrared excess (dust disk). \citet{2022ApJS..259...38Z} proposed an updated criterion to search for HAeBes and identified 71 HAeBes from LAMOST DR7. We used a similar method to search for F-type PMS stars.
The criterion on IR excess for HAeBes proposed by \citet{2022ApJS..259...38Z} is as follow.

\begin{enumerate}
\item HAeBes are located in the defined region on either color-color diagram.
\begin{enumerate}
\item ($\rm K_S$-W1,H-$\rm K_S$) diagram\\
$(H-K_S)_0 > 0.4$ and $(K_S-W1)_0 > 0.8$
\item (H-$\rm K_S$,J-H) diagram\\
$(J-H)_0 < 1.625 \times (H-K_S)_0 - 0.1$\\
$(J-H)_0 > 0.58 \times (H-K_S)_0 - 0.24$\\
$(J-H)_0 > -0.9 \times (H-K_S)_0 + 0.75$
\end{enumerate}

\item HAeBes are located in the defined region on the (W2-W3,W1-W2) diagram.\\
$(W2 - W3)_0 > 2.0$\\
$(W1 - W2)_0 > 0.25$\\
$(W1 - W2)_0 < 0.9 \times (W2 - W3)_0 - 0.25$\\
$(W1 - W2)_0 >-1.5 \times (W2 - W3)_0 + 2.1$

\item The target is an explicit point source on the WISE images of all four bands.
\end{enumerate}

\subsection{Data acquisition and confirmation} \label{subsec:confirmation}

Our initial catalog consists of 2,538,636 low-resolution spectra of F-type stars classified by the LAMOST 1D pipeline. The primary method of LAMOST 1D pipeline is the Chi-square matching with the templates of  \citet{2014AJ....147..101W}. The pipeline is designed for batch processing of enormous spectra instead of a single target. Thus it is not surprising that some non-F type spectra may be wrongly classified as F-type stars by the 1D pipeline. The poorly recognized spectra may be earlier or later than the F type, or the source may be not even a star when the quality of the spectrum is quite terrible in some extreme cases. It should be noted that all spectra have not been checked before being applied to the IR criterion.

\subsubsection{Data acquisition} \label{subsubsec:data}

We retrieved the IR photometry from the AllWISE source catalog \citep{AllWISE2013}. As part of the online point source catalog, a match to the 2MASS point source catalog \citep{2003yCat.2246....0C,2MASS.2006AJ....131.1163S} is automatically included, and we retrieved these data as well. To exclude the possible unreliable photometry, we adopted the criteria of photometry quality recommended by \citet{2003yCat.2246....0C,AllWISE2013,2016ApJS..224...23X}. In total, we got 1,965,114 entries with $J < 15.8$, $H < 15.1$, $K_S < 14.3$, $W1 < 17.1$, and $W2 < 15.7$ mag.

\subsubsection{Candidates} \label{subsubsec:candidates}

Extinction mainly consists of two components, interstellar extinction, and circumstellar extinction. Extinction will make a star appear redder and extinction correction will reveal its bluer nature. To include as many PMS candidates as possible for the further investigation and avoid the unnecessary uncertainty of extinction correction, no extinction correction has been taken before the candidate searching.

\begin{figure}[ht!]
\includegraphics[width=0.5\textwidth]{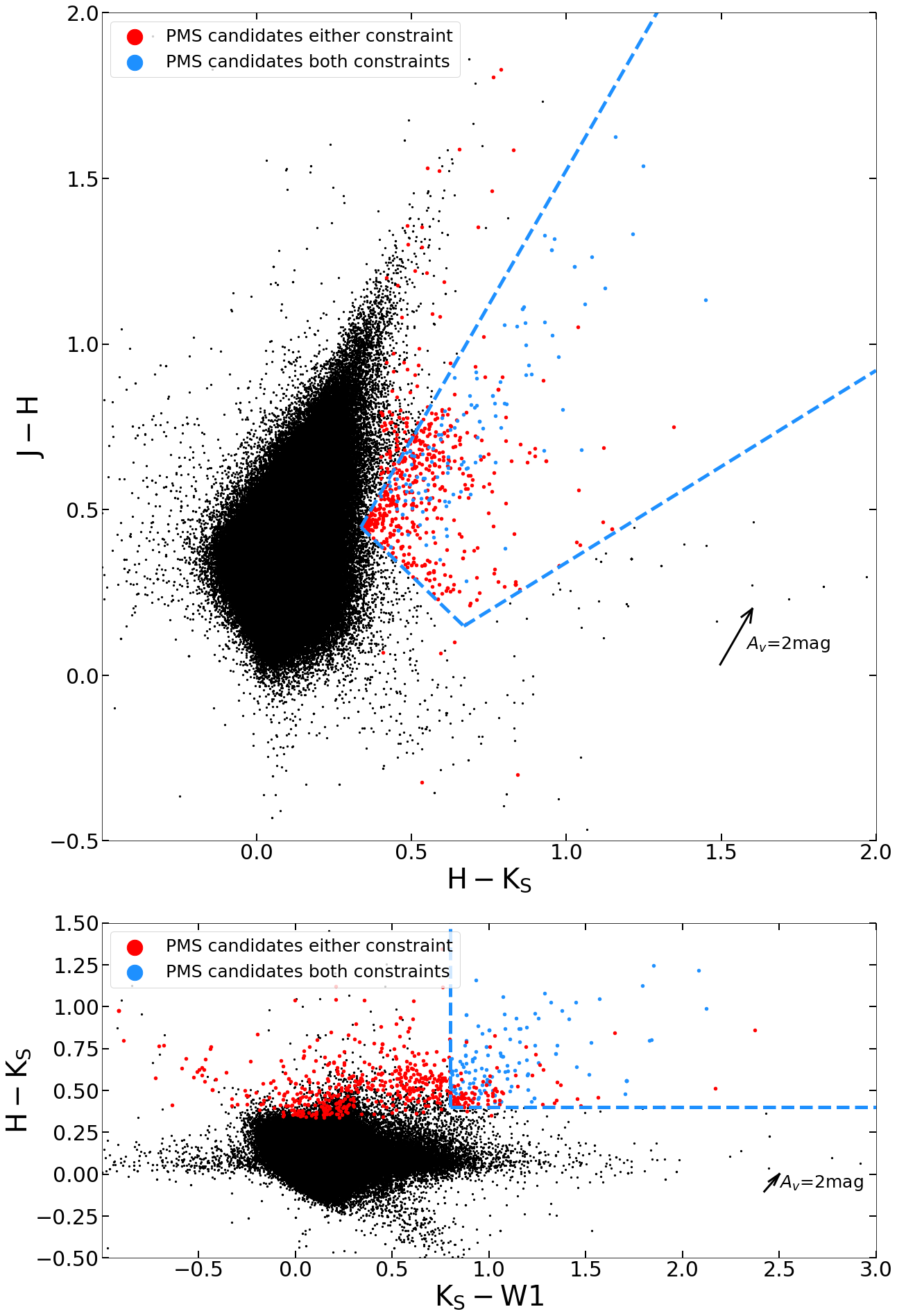}
\caption{The result of the first restriction, namely (H-$\rm K_S$,J-H) and ($\rm K_S$-W1,H-$\rm K_S$) diagram. Blue and red dots are the pre-main-sequence candidates. Blue ones are the common candidates through two diagrams, while red ones are only recognized as candidates in either diagram. The black dots represent the sources ruled out according to the first restriction.
\label{fig:fig1}}
\end{figure}

We applied the first restriction based on the (H-$\rm K_S$,J-H) and ($\rm K_S$-W1,H-$\rm K_S$) color-color diagrams to the whole entries. 712 PMS candidates were retrieved. As shown in Figure \ref{fig:fig1}, 139 candidates located in the defined region on both two color-color diagrams are marked with blue dots, and 573 candidates located in the defined region on either diagram are marked with red dots. Black dots represent the sources ruled out according to the first restriction.

\begin{figure}[ht!]
\includegraphics[width=0.5\textwidth]{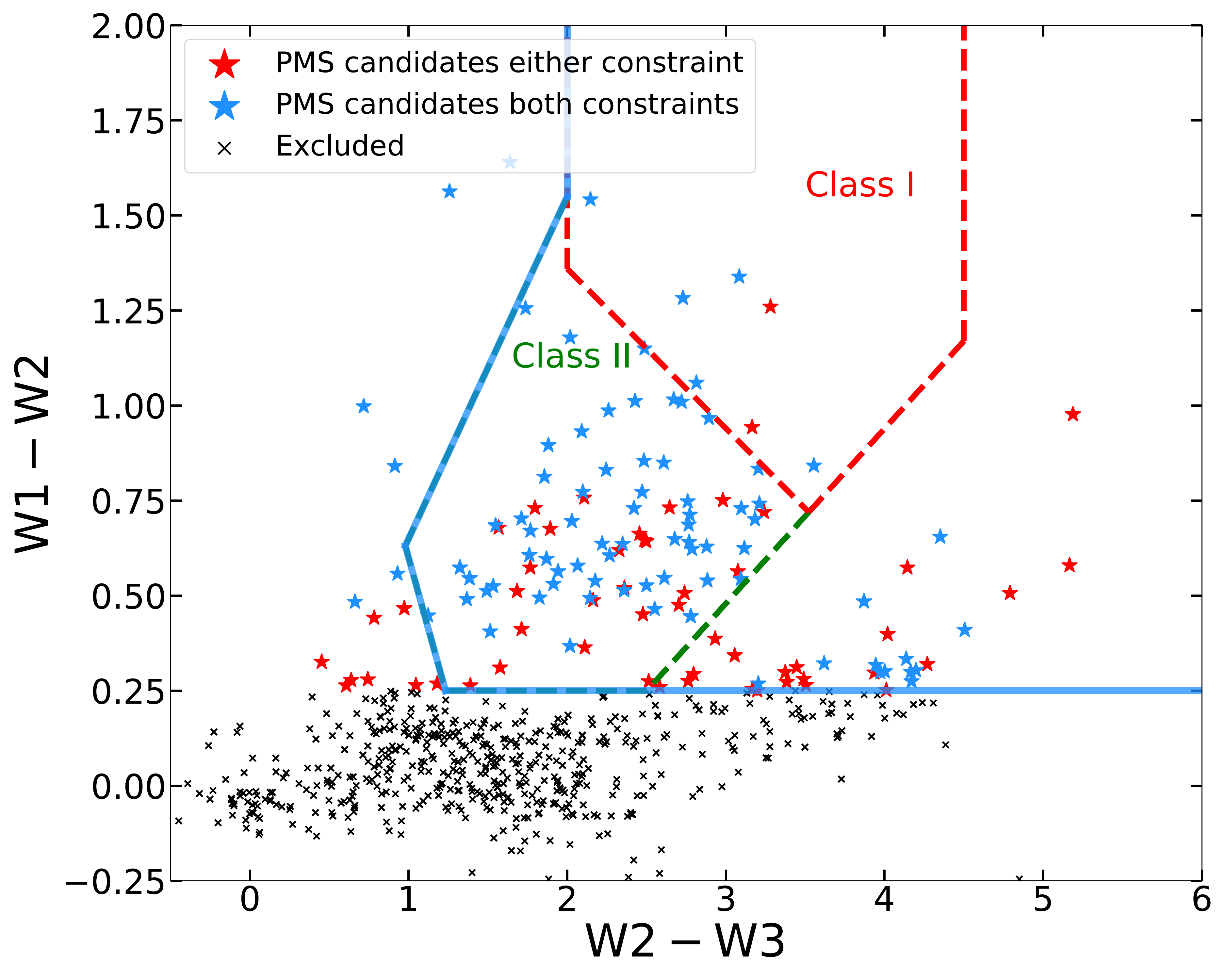}
\caption{The result of the second restriction, namely (W2-W3,W1-W2) diagram. The star marks represent the pre-main-sequence candidates. The cross marks are the sources excluded according to the second restriction. All the star and cross marks are from the blue and red dots in Figure \ref{fig:fig1}. The regions of Class \Rmnum{1} and \Rmnum{2} YSOs proposed by \cite{YSO1.2014ApJ...791..131K} are marked with red and green lines, respectively. The blue lines represent the constraints of HAeBes defined by \citet{2022ApJS..259...38Z}.
\label{fig:fig2}}
\end{figure}

We applied the second restriction based on the (W2-W3,W1-W2) color-color diagram. As extinction correction had not been applied, more PMS candidates should be included for the further inspection. We took the only one constraint $(W1 - W2)_0 > 0.25$ instead of the four constraints in the second restriction shown as the blue lines in Figure \ref{fig:fig2}. Finally, 153 F-type PMS candidates were retrieved shown as the star marks in Figure \ref{fig:fig2}. The wide distribution of $(W2 - W3)$ values in the excluded sources may be the consequence of falsely detected W3 values. When referred to the WISE images, these sources are usually too faint to be discerned on W3 band. Uncorrected interstellar extinction may also contribute to the scatter slightly, but it is not likely to be the main factor.
%A modification has been taken when applied to the second restriction based on (W2-W3,W1-W2) diagram.

\subsubsection{Candidate refining and confirmation} \label{subsubsec:refining}

The sample of the 153 PMS candidates is undoubtedly contaminated by some fake sources, such as possible galaxies, quasi stellar objects, etc. Candidate refining consists of two procedures based on images and spectra, respectively.

The first step is the quality inspection of photometry based on the WISE images. As is recommended by \citet{YSO1.2014ApJ...791..131K}, visual inspection is the most precise way to discern possible fake sources. To eliminate the fake sources in the WISE photometric data during YSO searching in large data set, \citet{YSO1.2014ApJ...791..131K} provided an empirical photometric filtration scheme (uncertainty/signal-to-noise/chi-squared criteria) and \citet{V.2019MNRAS.487.2522M} suggested a Random Forest method, both of which are based on the result of the visual inspection. Considering the accuracy and our moderate data size, we visually inspected the WISE images of the 153 PMS candidates on all four bands.

The second step is visual inspection of optical spectra. We removed possible galaxies and picked out spectra with discernable photospheric continuum as the final candidates for the further confirmation.

We made the confirmation progress one by one based on the WISE images and optical spectra. When a source is a distinct point source on four bands of the WISE images and has good-quality spectra free of intense nebula emissions ([N\Rmnum{2}]6549, 6583$\rm\AA$, [S\Rmnum{2}]6717, 6731$\rm\AA$), [O\Rmnum{3}]4959, 5007$\rm\AA$), we confirm it as a PMS star (location on the HR diagram is shown in Figure \ref{fig:fig5}). When a source is affected by the environment (molecular clouds or bright nearby stars) on the WISE images, its spectra are contaminated by intense nebula emission lines or its spectra are greatly affected by noise, we confirm it as a PMS candidate. We got 21 PMS stars and 10 PMS candidates along with one evolved star J191512.12+394250.5, a well-known post-AGB binary (IRAS 19135+3937, \citet{2019A&A...631A..53B,2019A&A...629A..49O}).

We also checked the sources ruled out according to the first restriction but still meet the second restriction to avoid possible omissions. 12 additional PMS candidates were retrieved. All of the 12 candidates locate close to the selected region defined by the upper panel of Figure \ref{fig:fig1}.

\begin{figure*}
\begin{center}
\includegraphics[width=\textwidth]{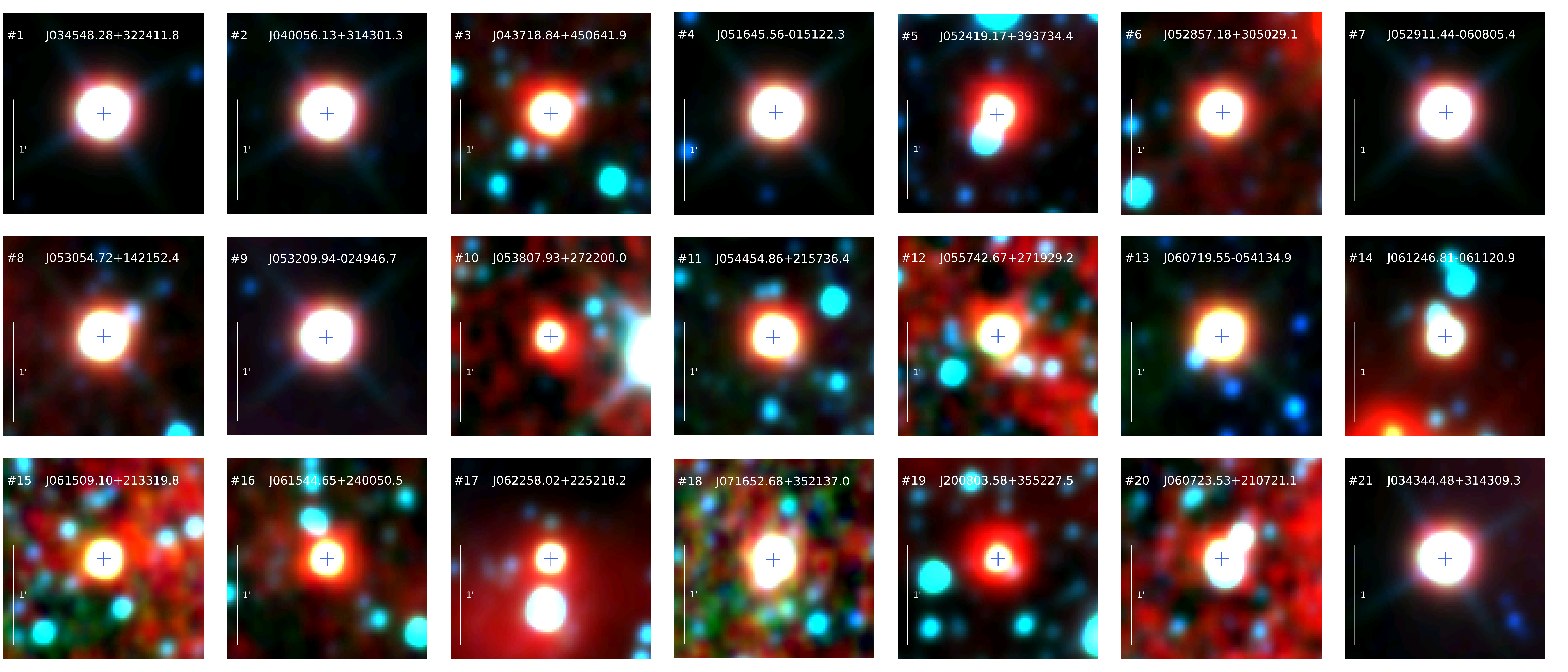}
\end{center}
\caption{WISE pseudo-color images (W1 in blue, W2 in green, and W4 in red) of our 20 F-type Herbig stars and \#21 (an outbursting EXor Herbig star, possible precursor of a Herbig Ae star). The target is marked with a blue cross in the middle of each image. All images are cut as 2'$\times$2'. The linear stretch range is from 5\% to 95\%.
\label{fig:fig3}}
\end{figure*}

Finally, we got 21 PMS stars and 22 PMS candidates in total. As shown in Figure \ref{fig:fig3}, we plotted the WISE pseudo-color images\footnote{\url{https://irsa.ipac.caltech.edu/applications/wise/}} (W1 in blue, W2 in green, and W4 in red) of the 21 PMS stars.

%20 F-type Herbig stars and one additional possible precursor of HAeBe \#21

We also measured the FWHM of H$\alpha$ emission lines. The average FWHM of 21 PMS stars is 7.43$\rm\AA$, corresponding to 339 km $\rm s^{-1}$. The average FWHM of 22 PMS candidates is 6.49 $\rm\AA$, corresponding to 296 km $\rm s^{-1}$. Besides, the 22 PMS candidates listed in Table \ref{tab:candidates} are not included in the following analysis. It should be noted that 21 PMS stars were retrieved, of which 20 are F-type Herbig stars as the main focus, and one is an outbursting EXor Herbig star discussed in Section \ref{subsec:EXor}.

{
\setlength\LTleft{-1in}
\setlength\LTright{-1in plus 1 fill}
\setlength{\tabcolsep}{5pt}

\small
\tiny
\begin{deluxetable*}{ccccccc}
\tablenum{1}
\tablecaption{F-type Herbig stars\label{tab:PMS1}}
\tablewidth{0pt}
\tablehead{
\colhead{No.} & \colhead{Designation} & \colhead{R.A. [degree]} & \colhead{Decl. [degree]} & \colhead{Specname} & \colhead{B [mag]} & \colhead{V [mag]}
}
\startdata
 1 & J034548.28+322411.8 & 56.45117 & 32.40329 & spec-58423-TD034419N321717S01\_sp04-055.fits & 13.363 & 12.05\\
 2 & J040056.13+314301.3 & 60.2338894 & 31.7170317 & spec-56981-HD040533N312949V01\_sp03-164.fits & 12.502 & 11.645\\
 3 & J043718.84+450641.9 & 69.3285 & 45.111639 & spec-57299-GAC067N44B1\_sp09-181.fits & 15.016 & 14.133\\
 4 & J051645.56-015122.3 & 79.189855 & -1.8562039 & spec-56966-HD052428S022349V01\_sp14-214.fits & 11.354 & 10.747\\
 5 & J052419.17+393734.4 & 81.079912 & 39.626228 & spec-57042-GAC084N40M1\_sp10-118.fits &   &   \\
 6 & J052857.18+305029.1 & 82.238266 & 30.841442 & spec-55910-GAC\_078N28\_B1\_sp12-049.fits & 16.037 & 14.702\\
   &                     &           &           & spec-55907-GAC\_079N29\_B1\_sp15-216.fits &   &   \\
 7 & J052911.44-060805.4 & 82.297671 & -6.134844 & spec-56617-VB081S05V2\_sp08-011.fits & 11.01 & 10.447\\
 8 & J053054.72+142152.4 & 82.7280228 & 14.3645656 & spec-56958-HD052834N134044V01\_sp09-154.fits & 15.343 & 14.403\\
 9 & J053209.94-024946.7 & 83.0414269 & -2.8296622 & spec-56966-HD052428S022349V01\_sp06-094.fits & 12.345 & 11.495\\
   &                     &            &            & spec-57083-HD053813S011009V01\_sp02-050.fits &        & \\
10 & J053807.93+272200.0 & 84.533079 & 27.366684 & spec-56218-GAC083N27B1\_sp04-114.fits & 16.881 & 15.493\\
11 & J054454.86+215736.4 & 86.228619 & 21.960137 & spec-56680-GAC085N22B1\_sp08-024.fits & 15.309 & 14.394\\
12 & J055742.67+271929.2 & 89.427826 & 27.324787 & spec-56350-GAC089N28V2\_sp01-150.fits & 15.55 & 14.529\\
13 & J060719.55-054134.9 & 91.831464 & -5.693037 & spec-56683-HD060658S061209B\_sp04-219.fits & 16.671 & 15.783\\
14 & J061246.81-061120.9 & 93.195058 & -6.189147 & spec-56671-VB092S06V2\_sp09-048.fits & 14.945 & 13.803\\
15 & J061509.10+213319.8 & 93.787935 & 21.555504 & spec-57753-GAC093N19M1\_sp12-133.fits &   &   \\
   &                     &           &           & spec-56983-GAC093N22M1\_sp08-207.fits &   &   \\
16 & J061544.65+240050.5 & 93.936066 & 24.014044 & spec-56609-GAC093N22M1\_sp09-057.fits &   &   \\
17 & J062258.02+225218.2 & 95.741764 & 22.871742 & spec-56658-GAC096N23B2\_sp05-057.fits &   &   \\
18 & J071652.68+352137.0 & 109.21954 & 35.360292 & spec-57104-GAC107N36B2\_sp07-191.fits & 15.014 & 14.535\\
19 & J200803.58+355227.5 & 302.0149308 & 35.8743108 & spec-57652-HD195955N370234B01\_sp06-214.fits & 15.367 & 14.426\\
20 & J060723.53+210721.1 & 91.848052 & 21.122555 & spec-56983-GAC093N22M1\_sp02-136.fits &   &
\enddata
\tablecomments{The `No.' column is the sequential identifier of our F-type Herbig stars. The `Specname' column is the FITS name of each spectrum. The B- and V- band photometry are from the APASS.
}
\end{deluxetable*}
}

\section{Stellar parameters} \label{sec:parameters}
%W2020
%Measurement of accretion rate and circumstellar disk nature requires pricise stellar parameters. Therefore,
This section aims to determine the stellar parameters, such as the effective temperature, distance, reddening, stellar luminosity, mass, age, and radius. The parameters were determined by combining the LAMOST spectra, stellar model atmosphere grids, photometry, Gaia EDR3 parallaxes, and pre-main-sequence evolutionary tracks.

\subsection{Effective temperature} \label{subsec:Teff}

A Chi-square match method was taken to determine the effective temperature ($\rm T_{eff}$). The stellar model referenced is the synthetic Castelli \& Kurucz model \footnote{\url{http://kurucz.harvard.edu/grids.html}} provided by \citet{ATLAS9.2003IAUS..210P.A20C}. The temperature grid we took is from 5600 K (typical temperature for G4) to 7500 K (typical temperature for A8) with the step of 100 K. The typical surface gravity log(g) of PMS stars is 3.5$-$4.5, and the value of log(g) between 3.5 and 4.5 does not have a significant effect on the spectra of the Castelli \& Kurucz model (compared in the same $\rm T_{eff}$). Thus, we adopted log(g)=4.0 for the model spectra. We also assumed solar metallicity and adopted the abundance [M/H]=0.0.

To determine the optimum wavelength range for stellar parameter estimation, we compared two different wavelength ranges, one with the full range of $\rm 3800-9000\AA$, the other with H$\alpha$ and H$\beta$ line masked which might be greatly influenced by emission lines (keeping $\rm 3800-4800\AA$, $\rm 4900-6500\AA$, and $\rm 6600-9000\AA$). Result turns out that the former is better by checking the fitting results with two ranges. We suggest that the wings of H$\alpha$ and H$\beta$ profiles may provide valuable information during model fitting despite the emission cores may bring in some interferences. The line wings of Balmer lines were used in previous HAeBes works. For example, \citet{W2020MNRAS.493..234W} determined the effective temperature and surface gravity of HAeBes by comparing the wings of the observed hydrogen Balmer lines with model spectra. Finally, the wavelength coverage adopted for fitting is $\rm 3800-9000\AA$. It should be noted that rotational velocity was not taken into account because the low resolution and inaccurate rotational velocities may also bring in larger uncertainties.

We reduced the high-resolution model spectra (R$\sim$50,000) into the same resolution as LAMOST (R$\sim$1800). We also manually checked and corrected the `redshift' z provided by LAMOST pipeline. We normalized the flux of both observed and synthetic spectra using the same method as follows.

\begin{equation}
F_i = \frac {f_i}{\sqrt{\frac{\Sigma f_i^2}{n}}}
\end{equation}
$F_i$ is the normalized flux and $f_i$ represents the flux before normalization. $n$ is the total number of sampling points. The denominator represents the root-mean-square of flux.

We used an empirical five-degree polynomial for pseudo-continuum normalization, and we compared the spectral lines between our observed spectra and synthetic spectra during the model fitting progress to avoid flux calibration error caused by LAMOST data reduction.
%Each observed spectrum was cross-matched with a grid, and the continuum was also corrected to each template with a five-degree polynomial during the paring process to avoid flux calibration error caused by LAMOST data reduction.

For each spectrum, we made the fitting process at each sampling point of the grid (5600 K$-$7500 K, 100 K for step) and calculated the corresponding chi-square value. Besides the quantitative Chi-square values, we also visually checked the matching results based on published spectral standards (Digital Spectral Classification Atlas by Gray \footnote{\url{http://ned.ipac.caltech.edu/level5/Gray/frames.html}}, \citet{2009ssc..book.....G}). In most cases, the optimal matching has the minimum Chi-square value. But when the quality of spectrum is not good enough such as seriously contaminated by noise, $\rm T_{eff}$ with minimum Chi-square value tends to be lower than the optimal matching. Noise strengthens the profile of metal lines and misleads the Chi-square value. In F-type spectra, metal lines enhance as temperature falls. Thus, visual inspection based on published spectral standard is also necessary.

The uncertainties of $\rm T_{eff}$ are set to be the step of grid, namely 100 K in most cases. For the spectra with low signal-noise ratio, the uncertainties are set to be double, namely 200 K. As shown in Figure \ref{fig:fig4}, we used the pseudo-continuum of the best-fitting Castelli \& Kurucz model to correct for the reddening along with the instrumental error, and plotted the optical spectra (blue) of 20 F-type Herbig stars, with the best-fitting Castelli \& Kurucz model (gray) superposed. It should be noted that the LAMOST spectra are joint with the blue and red arms processed by different CCDs. Thus, in some cases, such as \#6 and \#14, match on the joint place may not be good enough as others. Besides the joint place, the instrumental error is another cause. For example, the wave-like structure between $\rm7200 \AA$ and $\rm8000 \AA$ in the spectrum of \#19 is due to the instrumental error rather than the intrinsic features. But these issues have little influence on our general Chi-square fitting method using the spectral lines.

\begin{figure*}
\includegraphics[width=\textwidth]{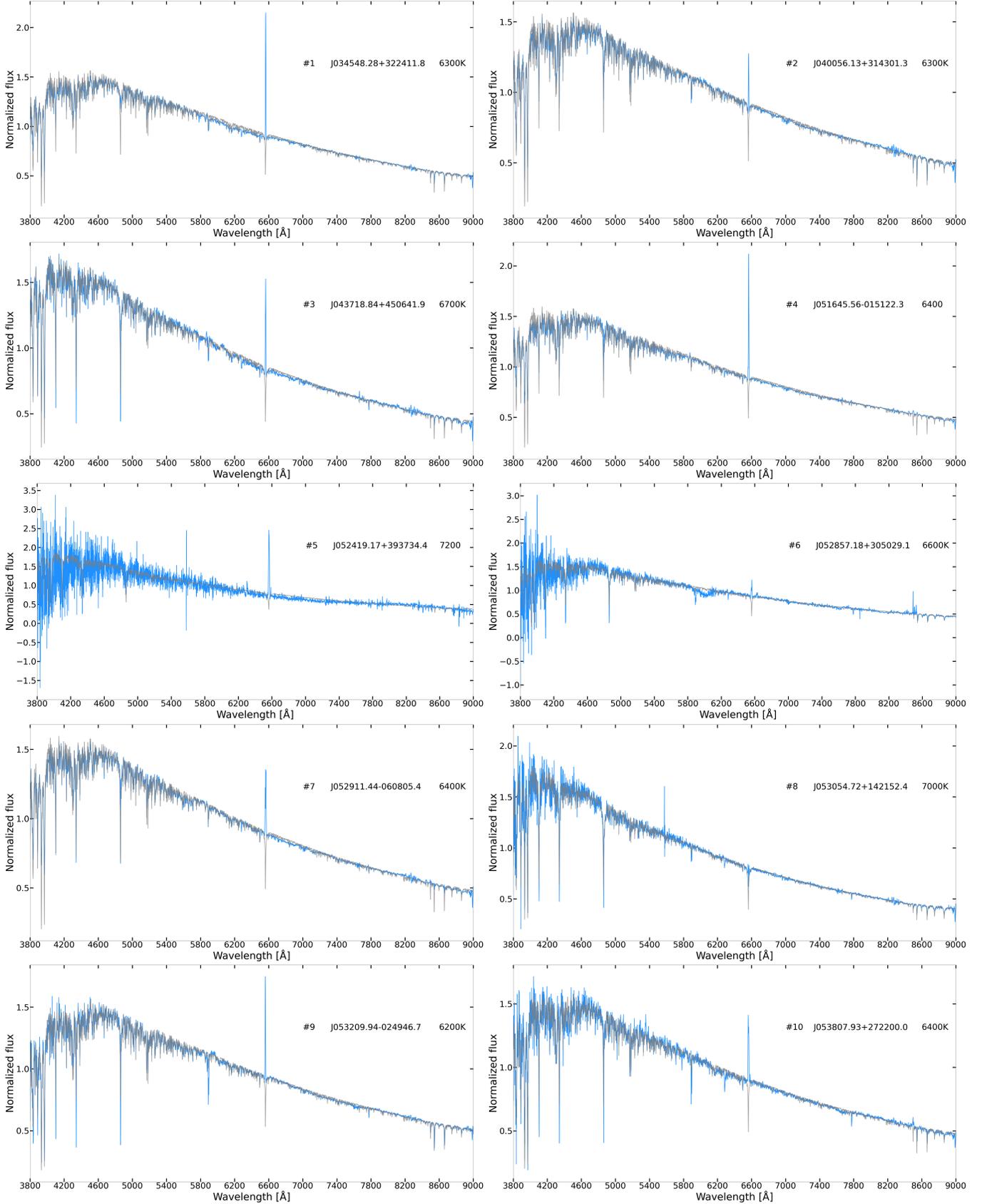}
\caption{Optical spectra (blue) of our 20 F-type Herbig stars, with the best-fitting Castelli \& Kurucz model (gray) superposed. The pseudo-continuum of the best-fitting Castelli \& Kurucz model has been used in the optical spectrum to correct for the reddening along with the instrumental error.
\label{fig:fig4}}
\end{figure*}

\addtocounter{figure}{-1}

\begin{figure*}
\includegraphics[width=\textwidth]{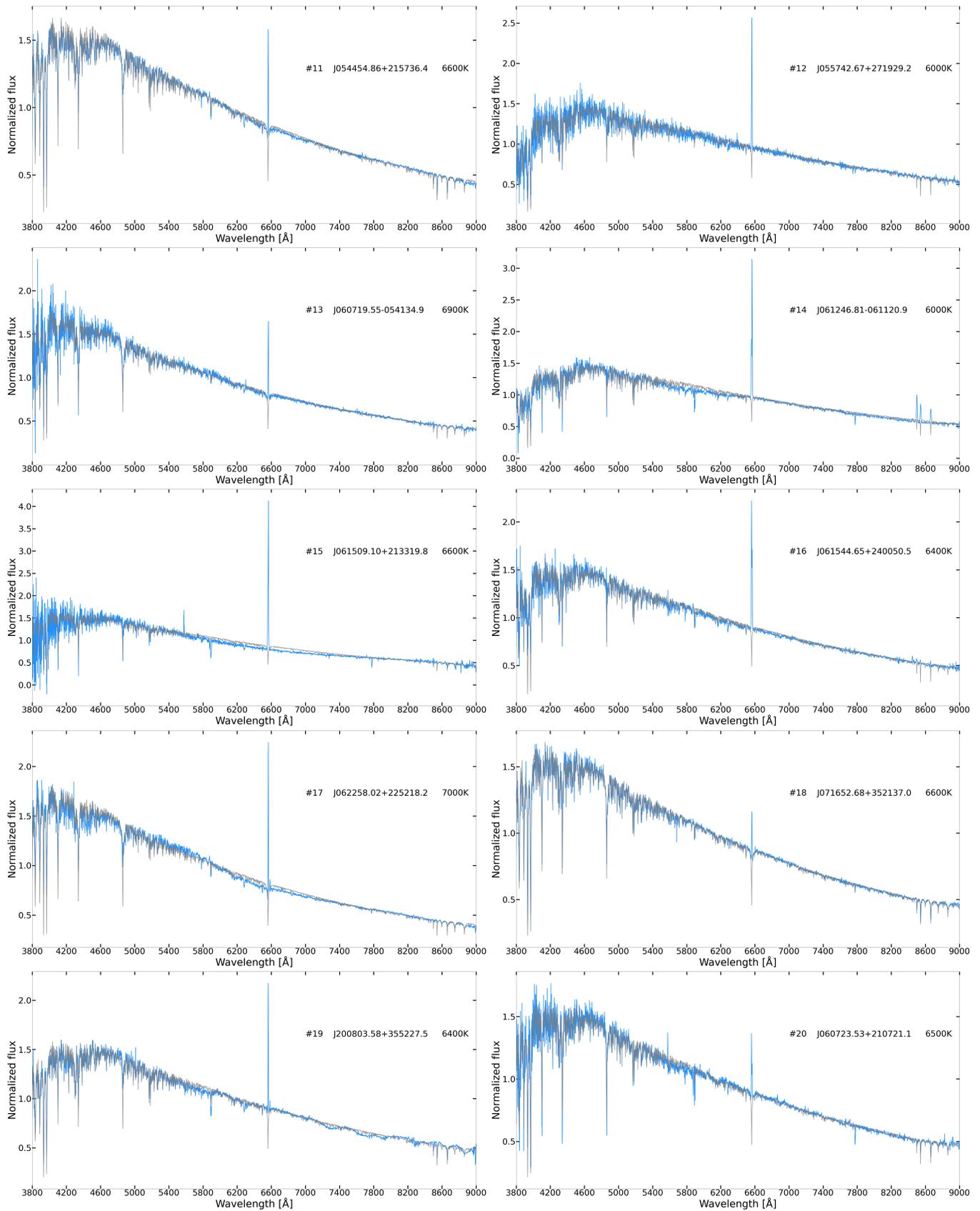}
\caption{continued.
\label{fig:fig4}}
\end{figure*}

\subsection{Distance} \label{subsec:distance}

To estimate extinction and stellar luminosity, and to make the further SED analysis, the distance information is essential. Gaia EDR3 provided full astrometric data for 1.468 billion sources, including positions, parallaxes and proper motions \citep{2021A&A...649A...2L}. We obtained the Gaia EDR3 parallax to each star. We estimated the distances by inversion of the parallaxes for 15 stars with $\sigma (\varpi) / \varpi \leq 0.1$. For five stars with $ 0.1 \leq \sigma (\varpi) / \varpi \leq 1.2$ (\#5, \#13, \#15, \#16, and \#20), we adopted the geometric distances from \citet{2021AJ....161..147B}.

\subsection{Extinction and Luminosity} \label{subsec:extinction}

We collected B- and V- band photometry for 15 stars from the AAVSO Photometric All Sky Survey (APASS). With the TOPCAT tool provided by \citet{TOPCAT2005ASPC..347...29T} and the VizieR, 14 of them were collected from APASS DR9 \citep{2016yCat.2336....0H} and a single one \#18 was collected from UCAC4 (\citet{2012yCat.1322....0Z}, gathered from APASS DR6). Five stars without the APASS photometry (\#5, \#15, \#16, \#17, and \#20) were managed with the SED fitting method described as the second part in this section.
%Two additional ones \#15 and \#16 are added with online searching of APASS DR10\footnote{\url{https://www.aavso.org/download-apass-data}}.

\subsubsection{15 stars with the APASS photometry} \label{subsec:APASS}

We obtained the intrinsic Johnson-Cousins color $(B-V)_{intrinsic}$ and V-band bolometric correction $BC_V$ for each spectral class from Table 6 of \citet{2013ApJS..208....9P}. E(B-V) was derived from the observed color $(B-V)_{observed}$ and the intrinsic one $(B-V)_{intrinsic}$. We adopted total-to-selective extinction $R_V=3.1$ for all stars (except \#13 with large IR excess, where $R_V=5$ was adopted as suggested by \citet{2004AJ....127.1682H}, Section \ref{subsec:SED} for details). We derived the absolute magnitude $M_V$ with V-band photometry, $A_V$, and the distance $d$. Then we derived the absolute bolometric magnitude $M_{bol}$ according to $M_{bol}=M_V+BC_V$. With the absolute bolometric magnitude of the Sun $M_{bol,\odot}=+4.74$, stellar luminosity $L$ was derived in the unit of the solar luminosity $L_\odot$. The uncertainty of stellar luminosity was derived from the uncertainties of $\rm T_{eff}$, V-band photometry, and distance.

\begin{equation}
M_V = V - A_V - 5log_{10}(\frac{d}{10pc})
\end{equation}

\begin{equation}
log_{10}{\frac{L}{L_\odot}}=\frac{M_{bol,\odot}-M_{bol}}{2.5}
\end{equation}

\subsubsection{5 stars analyzed with the SED fitting method} \label{subsec:SED fitting}

Five stars without the APASS photometry were analyzed with the SED fitting method assisted by the online tool Virtual Observatory SED Analyser (VOSA)\footnote{\url{http://svo2.cab.inta-csic.es/theory/vosa/}}, developed by the Spanish Virtual Observatory. \citet{2021A&A...650A.182G} made a study of 209 HAeBes gathered from literature, based on the results of VOSA. It should be noted that the best-fit output of VOSA is highly relevant to the input ranges of parameters, especially the $\rm T_{eff}$, distance, and extinction. In most cases, the three parameters above always degenerate with each other. Benefiting from the known $\rm T_{eff}$ range from LAMOST spectra and the known distance range from Gaia EDR3 parallaxes, the VOSA output $A_V$ and $L$ are convincing.

We included photometric data from Gaia EDR3 \citep{2021A&A...649A...3R}, 2MASS \citep{2MASS.2006AJ....131.1163S}, and WISE \citep{WISE.2010AJ....140.1868W}. According to \citet{2021A&A...650A.182G}, photometry at wavelength shorter than U band or longer than J band should not be used to fit the photospheres for the possible excess due to the accretion and dust emission. Considering our $\rm T_{eff}$ uncertainties are 100 or 200 K and the grid step of the model is 250 K, Gaia EDR3 photometry alone ($G_{BP}, G, G_{RP}$) is adequate for fitting.

We adopted the same atmosphere model as mentioned in Section \ref{subsec:Teff} for the SED fitting, namely Castelli \& Kurucz model. The values of log(g) and metallicity have little effect on the SED shape \citep{W2020MNRAS.493..234W,2021A&A...650A.182G,2022ApJ...930...39V}. We assumed the input model parameters $log(g)=4.0$ and $[M/H]=0.0$. We adopted a Chi-square fit method and used the Chi-square instead of the reduced Chi-square. The best-fit $A_V$ along with the stellar luminosity and uncertainty can be obtained as the output of VOSA. Uncertainty of the stellar luminosity provided by VOSA is supposed to be mainly attributed to the input uncertainty of distance.

{
\setlength\LTleft{-1in}
\setlength\LTright{-1in plus 1 fill}
\setlength{\tabcolsep}{5pt}

\small
\tiny
\begin{deluxetable*}{cccccccccc}
\tablenum{2}
\tablecaption{Parameters of F-type Herbig stars\label{tab:PMS2}}
\tablewidth{0pt}
\tablehead{
\colhead{No.} & \colhead{Designation} & \colhead{$\rm T_{eff}$ [K]} & \colhead{SpType} & \colhead{Distance [pc]} & \colhead{$\rm A_V$} & \colhead{$\rm log(L/L_\odot)$} & \colhead{Mass [$\rm M_\odot$]} & \colhead{Age [Myr]} & \colhead{Radius [$\rm R_\odot$]}
}
\startdata
 1 & J034548.28+322411.8 & $6300^{+100}_{-100}$ & F6 & $318.2^{+3.5}_{-3.5}$ & $2.52^{+0.09}_{-0.09}$ & $1.11^{+0.06}_{-0.07}$ & $1.95^{+0.11}_{-0.10}$ & $5.54^{+0.82}_{-0.72}$ & $3.00^{+0.34}_{-0.31}$\\
 2 & J040056.13+314301.3 & $6300^{+100}_{-100}$ & F6 & $370.0^{+3.7}_{-3.6}$ & $1.11^{+0.09}_{-0.09}$ & $0.83^{+0.06}_{-0.07}$ & $1.58^{+0.08}_{-0.08}$ & $9.43^{+1.47}_{-1.08}$ & $2.19^{+0.24}_{-0.23}$\\
 3 & J043718.84+450641.9 & $6700^{+100}_{-100}$ & F2 & $1156.2^{+26.3}_{-25.1}$ & $1.56^{+0}_{-0.16}$ & $1.00^{+0.05}_{-0.11}$ & $1.68^{+0.06}_{-0.11}$ & $8.60^{+1.44}_{-0.66}$ & $2.34^{+0.21}_{-0.34}$\\
 4 & J051645.56-015122.3 & $6400^{+100}_{-100}$ & F5 & $367.9^{+1.8}_{-2.8}$ & $0.42^{+0}_{-0.09}$ & $0.91^{+0.02}_{-0.05}$ & $1.63^{+0.03}_{-0.05}$ & $8.89^{+0.78}_{-0.65}$ & $2.31^{+0.12}_{-0.19}$\\
 5 & J052419.17+393734.4 & $7200^{+200}_{-200}$ & F0 & $6049.8^{+2348.1}_{-1674.4}$ & 3.50 & $1.63^{+0.22}_{-0.48}$ & $2.65^{+0.54}_{-0.85}$ & $2.78^{+4.84}_{-1.08}$ & $4.21^{+1.54}_{-1.90}$\\
 6 & J052857.18+305029.1 & $6600^{+100}_{-100}$ & F4 & $1646.9^{+90.7}_{-81.7}$ & $2.81^{+0.16}_{-0.12}$ & $1.57^{+0.17}_{-0.15}$ & $2.72^{+0.41}_{-0.33}$ & $2.45^{+0.98}_{-0.77}$ & $4.69^{+1.16}_{-0.84}$\\
 7 & J052911.44-060805.4 & $6400^{+100}_{-100}$ & F5 & $355.8^{+4.1}_{-4.0}$ & $0.29^{+0}_{-0.09}$ & $0.94^{+0.03}_{-0.06}$ & $1.68^{+0.04}_{-0.08}$ & $8.28^{+1.07}_{-0.59}$ & $2.41^{+0.16}_{-0.23}$\\
 8 & J053054.72+142152.4 & $7000^{+100}_{-100}$ & F1 & $751.2^{+10.8}_{-10.5}$ & $1.86^{+0}_{-0}$ & $0.63^{+0.06}_{-0.06}$ & $1.46^{+0.04}_{-0.04}$ & $14.62^{+5.44}_{-0.01}$ & $1.40^{+0.15}_{-0.13}$\\
 9 & J053209.94-024946.7 & $6200^{+100}_{-100}$ & F7 & $350.5^{+2.5}_{-2.4}$ & $0.99^{+0.09}_{-0.06}$ & $0.80^{+0.10}_{-0.09}$ & $1.58^{+0.13}_{-0.10}$ & $9.19^{+1.65}_{-1.74}$ & $2.19^{+0.36}_{-0.28}$\\
10 & J053807.93+272200.0 & $6400^{+100}_{-100}$ & F5 & $2285.7^{+212.4}_{-179.1}$ & $2.84^{+0}_{-0.09}$ & $1.56^{+0.08}_{-0.10}$ & $2.77^{+0.19}_{-0.22}$ & $2.27^{+0.54}_{-0.36}$ & $4.91^{+0.63}_{-0.67}$\\
11 & J054454.86+215736.4 & $6600^{+100}_{-100}$ & F4 & $1360.7^{+42.2}_{-39.7}$ & $1.50^{+0.16}_{-0.12}$ & $1.01^{+0.13}_{-0.11}$ & $1.71^{+0.18}_{-0.12}$ & $8.23^{+1.57}_{-1.89}$ & $2.45^{+0.48}_{-0.36}$\\
12 & J055742.67+271929.2 & $6000^{+200}_{-200}$ & F9 & $1720.3^{+103.5}_{-92.4}$ & $1.43^{+0.09}_{-0.12}$ & $1.15^{+0.11}_{-0.11}$ & $2.22^{+0.18}_{-0.18}$ & $3.62^{+0.93}_{-0.89}$ & $3.48^{+0.73}_{-0.60}$\\
13 & J060719.55-054134.9 & $6900^{+200}_{-200}$ & F1 & $845.5^{+199.0}_{-131.6}$ & $2.74^{+0}_{-0.20}$ & $0.53^{+0.21}_{-0.16}$ & $1.42^{+0.09}_{-0.09}$  & $16.89^{+6.24}_{-3.89}$  & $1.29^{+0.44}_{-0.28}$\\
14 & J061246.81-061120.9 & $6000^{+200}_{-200}$ & F9 & $749.3^{+48.5}_{-42.9}$ & $1.80^{+0.09}_{-0.12}$ & $0.87^{+0.12}_{-0.12}$ & $1.77^{+0.18}_{-0.16}$ & $6.40^{+1.78}_{-1.68}$ & $2.52^{+0.58}_{-0.47}$\\
15 & J061509.10+213319.8 & $6600^{+200}_{-200}$ & F4 & $1803.1^{+197.5}_{-162.4}$ & $3.00$ & $0.72^{+0.08}_{-0.10}$ & $1.45^{+0.04}_{-0.06}$ & $13.54^{+1.80}_{-2.06}$ & $1.75^{+0.29}_{-0.28}$\\
16 & J061544.65+240050.5 & $6400^{+100}_{-100}$ & F5 & $1920.3^{+1151.7}_{-681.9}$ & $3.00$ & $0.87^{+0.23}_{-0.54}$ & $1.59^{+0.30}_{-0.37}$ & $9.47^{+11.87}_{-3.36}$ & $2.21^{+0.78}_{-1.06}$\\
17 & J062258.02+225218.2 & $7000^{+100}_{-100}$ & F1 & $1649.6^{+65.1}_{-60.3}$ & 3.00 & $1.53^{+0.03}_{-0.03}$ & $2.48^{+0.07}_{-0.09}$ & $3.24^{+0.34}_{-0.22}$ & $3.95^{+0.27}_{-0.26}$\\
18 & J071652.68+352137.0 & $6600^{+100}_{-100}$ & F4 & $2479.5^{+141.0}_{-126.6}$ & $0.15^{+0.16}_{-0.12}$ & $1.11^{+0.11}_{-0.09}$ & $1.84^{+0.16}_{-0.12}$ & $6.77^{+1.37}_{-1.25}$ & $2.73^{+0.46}_{-0.34}$\\
19 & J200803.58+355227.5 & $6400^{+100}_{-100}$ & F5 & $2207.0^{+95.0}_{-87.5}$ & $1.46^{+0}_{-0.09}$ & $1.41^{+0.06}_{-0.09}$ & $2.43^{+0.12}_{-0.16}$ & $3.18^{+0.63}_{-0.37}$ & $4.10^{+0.42}_{-0.50}$\\
20 & J060723.53+210721.1 & $6500^{+100}_{-100}$ & F5 & $2331.7^{+523.9}_{-454.6}$ & 2.00 & $0.69^{+0.15}_{-0.24}$ & $1.40^{+0.15}_{-0.09}$ & $13.70^{+4.31}_{-3.32}$ & $1.75^{+0.40}_{-0.46}$
\enddata
\tablecomments{The `No.' column is the sequential identifier of our F-type Herbig stars. The `$\rm T_{eff}$' column represents the stellar effective temperature derived from the best-fitting with Castelli \& Kurucz model \citep{ATLAS9.2003IAUS..210P.A20C}. The `SpType' column is the spectral type of LAMOST low-resolution spectra derived from the corresponding $\rm T_{eff}$. The `Distance' column is the distance to the Sun, derived from the parallax from Gaia EDR3. The extinction $\rm A_V$, stellar luminosity, mass, age, and radius columns are the determined results in this work.
}
\end{deluxetable*}
}

\subsection{Stellar Mass, Age, and Radius} \label{subsec:mass}

We used Hertzsprung-Russell (HR) diagram and pre-main-sequence evolutionary tracks to determine the stellar masses and ages for our PMS stars. Several famous evolutionary models were computed with different stellar evolutionary codes. The PARSEC \citep{2012MNRAS.427..127B,2017ApJ...835...77M} tracks and isochrones were computed with the PAdova and TRieste stellar evolution code. The Pisa \citep{2011A&A...533A.109T} PMS tracks and isochrones used the FRANEC evolutionary code. Despite slight differences, the well-tested stellar evolutionary models share the similar result. We adopted the latter one (the Pisa model, dedicated to PMS evolution) to estimate stellar masses and ages for our PMS stars. We adopted the solar chemical composition proposed by \citet{2011A&A...533A.109T}, namely the metallicity $Z=0.01377$, the initial helium abundance $Y=0.2533$, the mixing length parameter $\alpha=1.68$, and the initial deuterium abundance $X_D=2.0 \times 10^{-5}$.

\begin{figure*}[ht!]
\includegraphics[width=\textwidth]{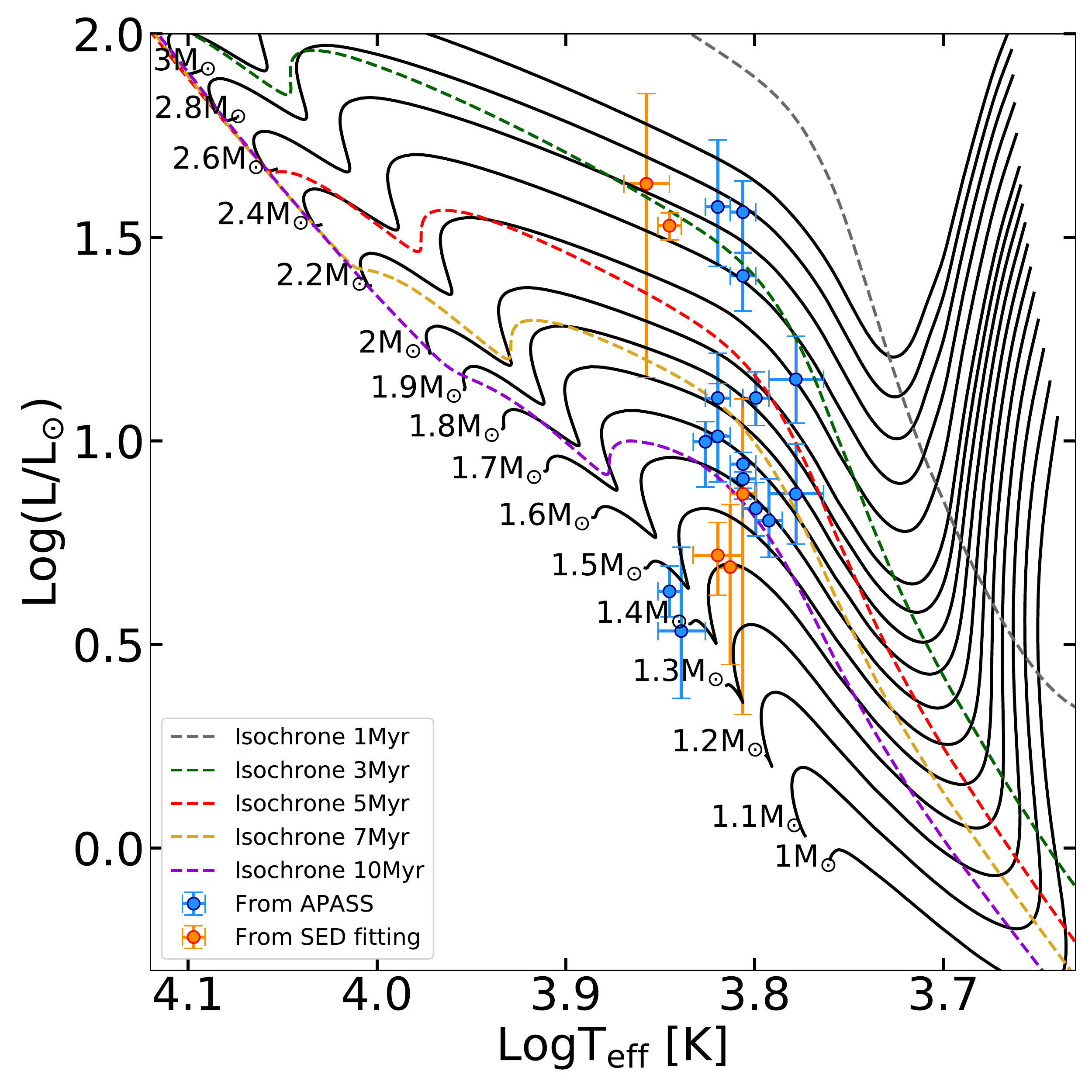}
\caption{HR diagram of the 20 F-type Herbig stars. 15 stars with the APASS photometry are marked with the blue dots. Five stars analyzed with the SED fitting method based on Castelli \& Kurucz model and Gaia EDR3 photometry are marked with the orange dots. The Pisa PMS evolutionary tracks and isochrones \citep{2011A&A...533A.109T} are also presented. The evolutionary tracks shown are limited up to $10^{1.5}$ Myr. Model parameters are settled as Metalicity Z=0.01377, Helium abundance Y=0.2533, Mixing length $\alpha$=1.68, and initial deuterium abundance $X_D=2.0 \times 10^{-5}$.
\label{fig:fig5}}
\end{figure*}

Using the interpolation with two closet points on the evolutionary tracks, stellar masses and ages were obtained. Uncertainties of masses and ages were derived from the uncertainties of $\rm T_{eff}$ and stellar luminosities. As shown in Figure \ref{fig:fig5}, the blue dots are the 15 stars with the APASS photometry, and the orange dots are the five stars analyzed with the SED fitting method based on Castelli \& Kurucz model and Gaia EDR3 photometry. The Pisa tracks ($1.0-3.0 M_\odot$) and isochrones (1,3,5,7,10 Myr) are also presented. The evolutionary tracks shown in Figure \ref{fig:fig5} are limited up to $10^{1.5}$ Myr.

It should be noted that pre-main-sequence evolutionary tracks partially overlap with post-main-sequence tracks. However, lacking dusty stellar winds, infrared excess is not expected for single intermediate-mass stars that have evolved to sub-giants \citep{2021A&A...652A.133V}.

16 stars with masses ranging from 1.58 to 2.77 $M_\odot$ were confirmed as F-type Herbig stars, 6 of which (\#5, \#6, \#10, \#12, \#17, and \#19) have masses larger than 2 $M_\odot$. Besides, \#8, \#13, \#15, and \#20 with masses about 1.5 $M_\odot$ were also classified as F-type Herbig stars in consideration of the uncertainties. In summary, we got 20 F-type Herbig stars in this work.
%\#8, \textbf{\#15,} and \#20 with initial mass less than 1.5 $M_\odot$ are confirmed as F-type T Tauri stars. \#13 is located closely below the main-sequence phase on the HR diagram and no mass or age has been measured for it. The only record in literature about \#13 is from \citet{2018AJ....156..241H} and \#13 is roughly classified as a long-period variable (LPV) whose variation is in some sense irregular (Section \ref{subsec:literature}). With relatively lower luminosity than others, we here classified \#13 as a potential F-type T Tauri star.

Using the Stefan-Bolzmann law, stellar luminosity can be expressed as $L=\sigma T^4 \times 4\pi R^2$. Thus, the stellar radius $R$ can be determined with the derived $\rm T_{eff}$ and stellar luminosity. The stellar luminosity and radius are both in the unit of the solar values.
%and the effective temperature of the Sun adopted is 5772 K.
\section{Analysis} \label{sec:analysis}

Detailed analysis of our F-type Herbig stars presented in this section are about SEDs, infrared spectra from Spitzer IRS enhanced product, and the records in literature.

\subsection{Spectral Energy Distributions} \label{subsec:SED}

\begin{figure*}
\begin{center}
\includegraphics[width=\textwidth]{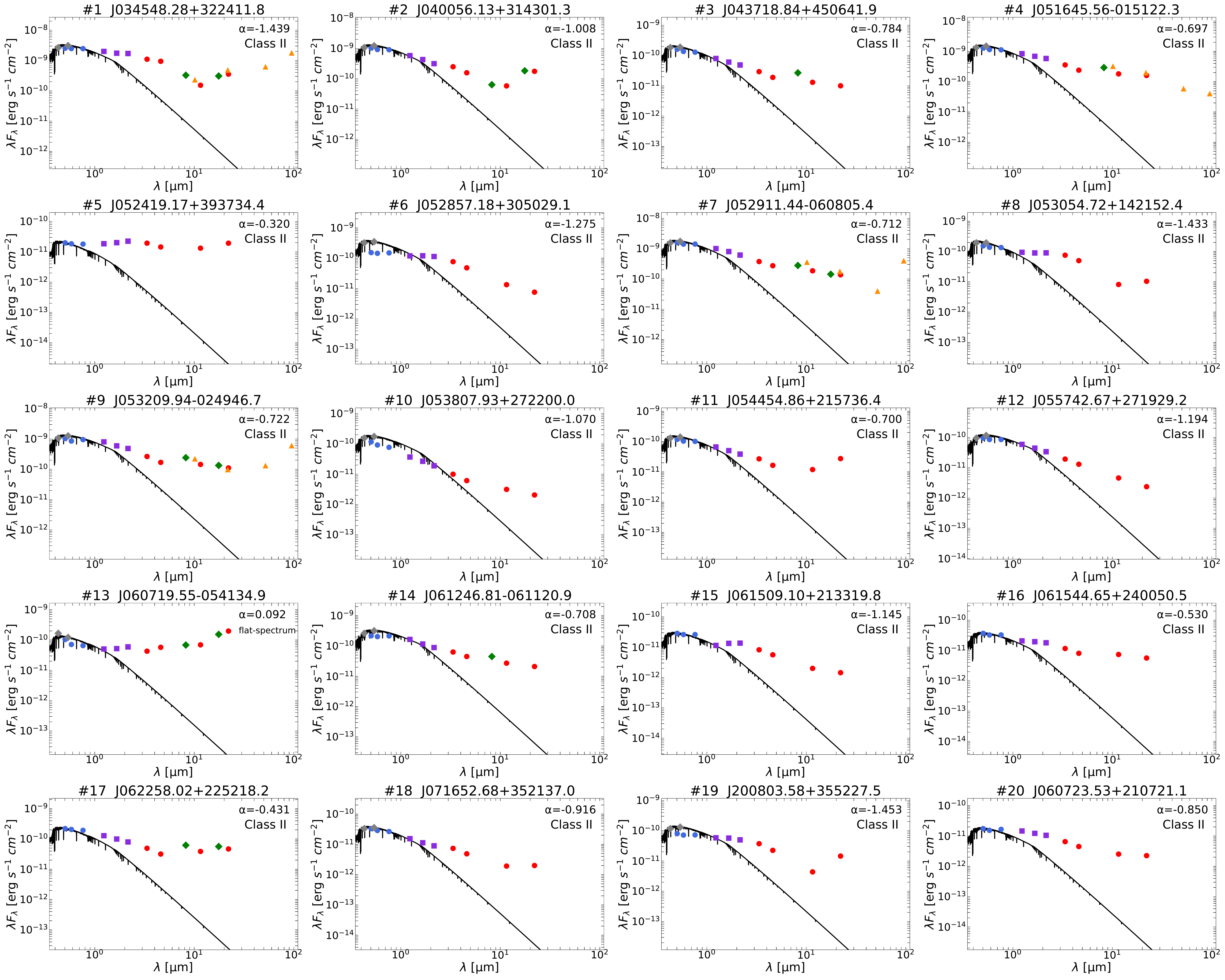}
\end{center}
%\plotone{percentage.pdf}
\caption{Spectral energy distributions of the 20 F-type Herbig stars. The photometric data included are from APASS DR9 (gray diamonds, corrected for reddening), Gaia EDR3 (blue dots, corrected for reddening), 2MASS (violet squares, corrected for reddening), WISE (red dots, corrected for reddening), IRAS (orange triangles), and AKARI (green diamonds). The Castelli \& Kurucz model spectrum with the same $\rm T_{eff}$ is in black. The slope $\alpha$ provided is the spectral index $\alpha_{K_S-W3}$.
\label{fig:fig6}}
\end{figure*}

The SEDs with Castelli \& Kurucz model are shown in Figure \ref{fig:fig6}. The photometric data included are from APASS DR9 (gray diamonds, corrected for reddening), Gaia EDR3 (blue dots, corrected for reddening), 2MASS (violet squares, corrected for reddening), WISE (red dots, corrected for reddening), IRAS (orange triangles, \citet{IRAS.1988iras....1.....B,IRAS.1988iras....7.....H}), and AKARI (green diamonds, \citet{AKARI.2010A&A...514A...1I}). The extinction law $A_\lambda/A_V$ adopted for reddening correction is from Table 3 of \citet{2019ApJ...877..116W}. The Castelli \& Kurucz model spectrum with corresponding $\rm T_{eff}$ is in black. In the SEDs of \#6 and \#10, the flux density of model spectra is little higher than Gaia EDR3 photometry. Possible variations are suggested to be the main cause. The extinctions and stellar luminosities of \#6 and \#10 were determined based on the APASS photometry rather than Gaia EDR3. Besides, as \citet{2021A&A...650A.182G,2009A&A...495..901M} pointed out, during the parameter determination of HAeBes, the photometry corresponding to the maximum of the stellar brightness (the star is likely to be less obscured) should be used in building the SEDs of variable stars.

The slope of SED can be used to discern the stage of YSO evolution. \citet{SED.1987IAUS..115....1L} first proposed the tripartite class system using the defined spectral index $\alpha = (\mathrm{d} log\lambda F_\lambda)/(\mathrm{d} log\lambda)$. $\lambda$ means the wavelength and $F_\lambda$ is the flux density at $\lambda$. The boundaries classified by \citeauthor{SED.1987IAUS..115....1L} are Class \Rmnum{1} with $0<\alpha \le 3$, Class \Rmnum{2} with $-2\le \alpha \le 0$, and Class \Rmnum{3} with $-3<\alpha \le -2$, and the slopes are determined between roughly 2 and 20 $\mu m$. Three categories represent three different evolutionary stages proposed by \citet{SED.1987IAUS..115....1L,1.2009ApJS..181..321E} (the envelope collapse as Class \Rmnum{1}, the accretion disk and star as Class \Rmnum{2}, and the disk dissipation during Class \Rmnum{3}). \citet{SED.1993ApJ...406..122A} added a Class 0 with high ratio of submillimeter to bolometric luminosity. \citet{SED.1994ApJ...434..614G} added a flat-spectrum class and formalized the modified system with the spectral index calculated between 2.2 and 10 $\mu m$. The modified 4-class system proposed by \citeauthor{SED.1994ApJ...434..614G} are detailed as follows. Class \Rmnum{1} has $\alpha \ge 0.3$, flat-spectrum class has $-0.3\le \alpha < 0.3$, Class \Rmnum{2} has $-1.6\le \alpha < -0.3$, and Class \Rmnum{3} has $\alpha < -1.6$. \citet{1994ApJ...434..330C} explained the flat spectrum as infall inducing, in which material fall onto a disk rather than the central star. They suggested that many flat-spectrum sources should be YSOs surrounded by dust infalling envelopes of substantial mass.

We adopted the similar band taken, $\rm K_S$ (2MASS, 2.159 $\mu m$) and W3 (WISE, 11.5608 $\mu m$), and determined the slopes of our Herbig stars.
\begin{equation}
\alpha = \rm \frac{log(\frac{\lambda_{W3} F_{W3}}{\lambda_{K_S} F_{K_S}})}{log(\frac{\lambda_{W3}}{\lambda_{K_S}})}
\end{equation}
$\lambda$ is the wavelength and $F_\lambda$ is the flux density at $\lambda$. According to the modified 4-class system proposed by \citet{SED.1994ApJ...434..614G}, 19 F-type Herbig stars belong to Class \Rmnum{2} YSOs, and the one \#13 ($\alpha=0.092$) belongs to the flat-spectrum class. \#5 ($\alpha=-0.320$) is a Class \Rmnum{2} YSO but it is near the boundary of the flat-spectrum class. The relatively large extinction and the young age of \#5 also agree with its early stage of disk evolution. However, \#13 is not the case. \#13 is bright on the IR bands and its derived age disagrees with its early stage of disk evolution indicated by the large IR excess. Despite taking a large value of total-to-selective extinction $R_V=5$ as suggested by \citet{2004AJ....127.1682H} for \#13 rather than the standard $R_V=3.1$, we might still underestimate its extinction and luminosity if the real $R_V$ was even larger. It should be noted that the quality flags of \#5 and \#13 from the AllWISE are all excellent, `AAAA' for ph\_qual (photometric quality of each band) and no saturation has been detected. Similar to \#5 and \#13, some other Herbig stars with large IR excesses are also presented by \citet{2021A&A...652A.133V,2021A&A...650A.182G,2001A&A...365..476M}. Statistic analysis of IR excesses can be seen as Section \ref{subsec:comparsion}. In general, the mean slope of the 20 F-type Herbig stars in our work is -0.865, and the result of SED analysis is quite consistent with their PMS nature.
%\citep{SED.1987IAUS..115....1L,SED.1994ApJ...434..614G,1994ApJ...434..330C} Another possible explanation is that \#13 is an close binary and its young companion with high IR excess is totally invisible on the optical band. But no direct evidence about its binary nature has been detected. We suggest an early stage of circumstellar disk evolution may account for them as well.

\subsection{Infrared Spectra} \label{subsec:infrared}

The silicate dust grains are the main components of the circumstellar disks and usually have the size of submicron in diameter. Silicate has two subtypes depending on the lattice structure, namely amorphous silicate and crystalline silicate. High temperatures are required to thermally anneal and crystallize amorphous grains. Crystalline silicate grains are considered as tracers of the history of protoplanetary, circumstellar disk \citep{2009A&A...507..327O}. Along with silicate, PAH molecules are also the tracers to probe the physical and chemical properties of protoplanetary disks (PAH molecules trace the gas disks). The infrared emission from PAH molecules is usually caused by stellar UV photons that excite electronic states in the PAH molecule, which subsequently de-excites through vibrational emission in stretching and bending mode resonances \citep{2021A&A...652A.133V}.

\begin{figure*}
\includegraphics[width=\textwidth]{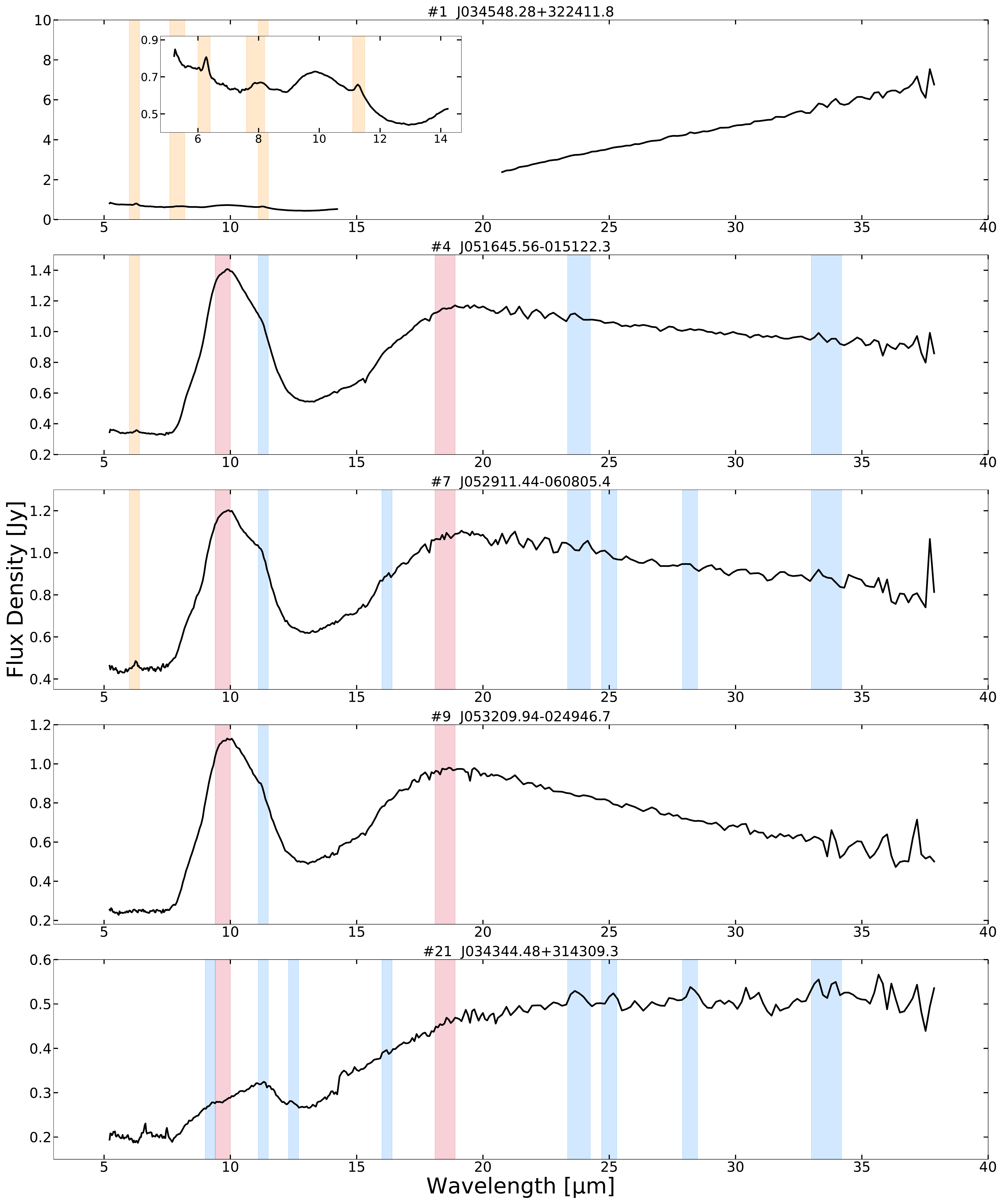}
\caption{Five Herbig stars with Spitzer IRS spectra. The PAH emission features at 6.2, 7.8, and 11.3 $\mu m$ are marked with light orange stripes. The amorphous silicate features peaking at 9.7 and 18.5 $\mu m$ are marked with light crimson stripes. The detected crystalline silicate features are marked with light blue stripes.
\label{fig:fig7}}
\end{figure*}

We obtained infrared spectra \footnote{\url{https://sha.ipac.caltech.edu/applications/Spitzer/SHA/}} for five stars (\#1, \#4, \#7, \#9, and \#21) from the Infrared Spectrograph (IRS, \citet{spitzer.2004ApJS..154...18H}) on the Spitzer Space Telescope \citep{spitzer.2004ApJS..154....1W}. The IRS spectra are shown in Figure \ref{fig:fig7} with PAH and silicate emission features marked with light-color stripes. We set the threshold for a detectable feature when the peak is 1.02 over the continuum strength. It should be noted that the detected crystalline silicate longer than 20 $\mu m$ may be disturbed by noise.

The infrared spectra of \#4, \#7, and \#9 are rather similar to each other, sharing the common features of amorphous silicate (peaking at 9.7 and 18.5 $\mu m$) and crystalline silicate (11.3 $\mu m$). Coincidentally, they also have similar parameters ($\rm T_{eff}$ 6200$-$6400 K, Mass 1.58$-$1.68 $M_\odot$, Age 8.28$-$9.19 Myr, Radius 2.19$-$2.41 $R_\odot$). Considering they are at the same evolutionary stage, similar infrared spectra are also reasonable.

Dislike three ones above, \#1 shows a rising infrared spectrum while \#21 shows a flat one on the mid-infrared band. Besides the rising continuum peaking farther than mid-infrared, the only detectable features of \#1 are the extremely weak PAH emissions at 6.2, 7.8, and 11.3 $\mu m$. The infrared spectrum of \#21 shows the evidence of silicate, both amorphous and crystalline. The detected crystalline silicate features at 9.2, 11.3, 12.5, 16.2, 23.8, 25, 28.2, and 33.6 $\mu m$ are marked with light blue stripes.

\subsection{Literature Records} \label{subsec:literature}

Using the SIMBAD, we searched records in literature for our F-type Herbig stars, and only several of them have the records. According to literature, five stars (\#1 EM* LkH$\rm \alpha$ 330, \#4 HD 290172, \#7 V* V1650 Ori, \#8 2MASS J05305472+1421524, and \#9 V* RY Ori) have been recognized as HAeBes or TTs, and two stars (\#2 IRAS 03577+3134 and \#14 2MASS J06124681-0611208) have been classified as YSOs.

\#1 is one of the most famous intermediate-mass T Tauri stars \citep{2021A&A...652A.133V,1.2018A&A...617A..83V,1.PAH2017ApJ...835..291S,1.2016A&A...590A..98H,1.2015ApJ...809..167B}. It was first classified as a T Tauri star with the spectral type of F6 by \citet{1.1977ApJ...214..747H}. Recently, \citet{1.spiral.2016AJ....152..222A,1.spiral.2018AJ....156...63U} detected a spiral-like structure of the \#1 disk suggesting a planet formation based on H-band linear polarimetric observations and H-, $\rm K_S$-bands observations separately. \citet{2021A&A...652A.133V} determined its parameters as 6240 K, 1.93 $M_\odot$, 4.66 Myr. In our work, we confirmed \#1 as an F-type Herbig star (6300 K, 1.95 $M_\odot$, 5.54 Myr).

\#4 is a known T Taur star which was first identified with a spectral type of G0 by \citet{1.1995AJ....109.2146T}. In our work, we confirmed \#4 as an F-type Herbig star (6400 K, 1.63 $M_\odot$, 8.89 Myr).

\#7 is a famous HAeBe star, and it was first identified as a PMS star by \citet{7.1990ApJS...74..575W}. It was classified as a HAeBe star or an IMTT star by various works \citep{2021A&A...652A.133V,7.2019A&A...622A..72V,7.2006A&A...457..171A,7.1994A&AS..104..315T,7.1992AJ....103..549G}. \citet{2021A&A...652A.133V} determined its parameters as 6160 K, 1.71 $M_\odot$, 6.20 Myr. In our work, we confirmed \#7 as an F-type Herbig star (6400 K, 1.68 $M_\odot$, 8.28 Myr).

\#8 was classified as a HAeBe star with a spectral type of F2 by \citet{8.2007ApJ...657..884L}. In our work, we confirmed \#8 as a an F-type Herbig star with the spectral type of F1 and the effective temperature of 7000 K. Considering its stellar mass (1.46 $M_\odot$) and uncertainty ($\pm0.04 M_\odot$), \#8 is on the boundary of Herbig stars.

\#9 is another famous HAeBe star and numerous studies have included it (6120 K, 1.69 $M_\odot$, 6.65 Myr in \citet{2021A&A...652A.133V}, 6250 K, 1.58 $M_\odot$, 9.42 Myr in \citet{2021A&A...650A.182G}, and 6250 K, 1.542 $M_\odot, 7.154 Myr$ in \citet{252HAeBe.2018A&A...620A.128V}). In our work, we confirmed \#9 as an F-type Herbig star (6200 K, 1.58 $M_\odot$, 9.19 Myr).
% The parameters provided by \citet{2021A&A...652A.133V} are 6120 K, 1.69 $M_\odot$, 6.65 Myr. The parameters determined by \citet{2021A&A...650A.182G} are 6250 K, 1.58 $M_\odot$, 9.42 Myr. The parameters calculated by \citet{252HAeBe.2018A&A...620A.128V} are 6250 K, 1.542 $M_\odot, 7.154 Myr$.

\#2 is a famous YSO, and it was first identified as a PMS star in an IRAS survey of the Taurus-Auriga molecular cloud by \citet{2.1990AJ.....99..869K}. It was also studied by \citet{2.1992ApJ...386..248B} with IRAS data. \citet{2.1996A&A...306..427C} further provided infrared CO absorption index for it. Recently, \citet{1.2015AJ....150...95A,2.2011ApJS..196....4R} made the studies to search for YSOs based on the WISE data, and \# 2 was also included in their work. In our work, we further identified \#2 as an F-type Herbig star with 1.58 $M_\odot$.

\#14 is a Class \Rmnum{2} YSO which was described as a PMS star with an optically thick disk \citep{14.2009ApJS..184...18G}. In our work, we further identified \#14 as an F-type Herbig star with 1.77 $M_\odot$.

\citet{2018AJ....156..241H} provided the first catalog of variable stars measured by Asteroid Terrestrial-impact Last Alert System (ATLAS), including 427,000 probable variables. According to their work, five stars (\#8 ATO J082.7280+14.3645, \#13 ATO J091.8314-05.6930, \#14 ATO J093.1950-06.1891, \#15 ATO J093.7879+21.5555, and \#16 ATO J093.9360+24.0140) were roughly classified as long-period variables (LPV) whose variation is in some sense irregular. It is worth mentioning that the only record in literature for \#13, \#15, and\#16 is the variability nature.

\subsection{Comparison with the LAMOST AFGK Catalog} \label{subsec:AFGK}

We cross-matched our F-type Herbig stars with the online LAMOST LRS Stellar Parameter Catalog of A, F, G, and K stars (AFGK catalog), in which 16 of the F-type Herbig stars have stellar parameters $\rm T_{eff}$, log(g), and [M/H]. In general, the provided effective temperatures agree with our values. Among the 16 F-type Herbig stars in the AFGK catalog, the mean value of log(g) (ranges from 3.744 to 4.374) is 4.01 dex, and the mean value of [M/H] is -0.05 dex. The provided values of log(g) and [M/H] are quite consistent with our assumptions, namely log(g)=4.0 and the solar metallicity.

\section{Discussion and Summary} \label{sec:discussion}

\subsection{An outbursting EXor Herbig star} \label{subsec:EXor}

Besides the 20 F-type Herbig stars, we also detected an outbursting EXor Herbig star \#21 J034344.48+314309.3 (R.A. 55.935352, Decl. 31.719259, IRAS F03406+3133, 2MASS J03434449+3143092), possible precursor of a Herbig Ae star.

 \#21 was first detected by \citet {1.2003AJ....126.1423K} in the observations of star-forming regions with the Midcourse Space Experiment. Then it was identified as a YSO candidate in the Perseus Molecular Cloud by \citet{1.2013ApJS..205....5H,1.2009ApJS..181..321E} using the data from the c2d Spitzer Legacy project. \citet{1.2014ApJ...782...51A} suggested \#21 is a bona fide young eruptive star (EXor) in the comparison study between the Spitzer and WISE data. They classified \#21 as a flat-spectrum class protostar, and provided the stellar parameters ($\rm T_{eff}$ as 5718--5209 K, stellar mass as 2.1--1.5 $M_\odot$, stellar luminosity as 13--5.4 $L_\odot$, the first value refers to the best fit and the second to the model with a 20\% higher chi-square). However, the stellar parameters of \#21 are still in controversy. \citet{1.2014ApJ...794..125C} provided a peculiar value (868 K) as the $\rm T_{eff}$ for \#21 with the forward modeling approach based on the spectra of the Apache Point Galactic Evolution Experiment (APOGEE). Also using the high-resolution H-band spectra of APOGEE, \citet{1.2015AJ....149....7C} identified it as a Be star (ABE-149) with H\Rmnum{1} Brackett series emission along with emissions of Fe\Rmnum{2} 16878, C\Rmnum{1} 16895, Mg\Rmnum{1} 15753 and 15770. However, \citet{1.2015AJ....150...95A} used a Bayesian Monte Carlo SED fitting method and estimated the stellar mass and age for \#21 as 0.8 $M_\odot$ and 1.05 Myr. Subsequently, \citet{1.2017AJ....153..174C} excluded \#21 from their classical Be study and agreed with the young nature proposed by \citet{1.2015AJ....150...95A}. No available optical spectra for \#21 before the LAMOST.

\begin{figure*}
\includegraphics[width=\textwidth]{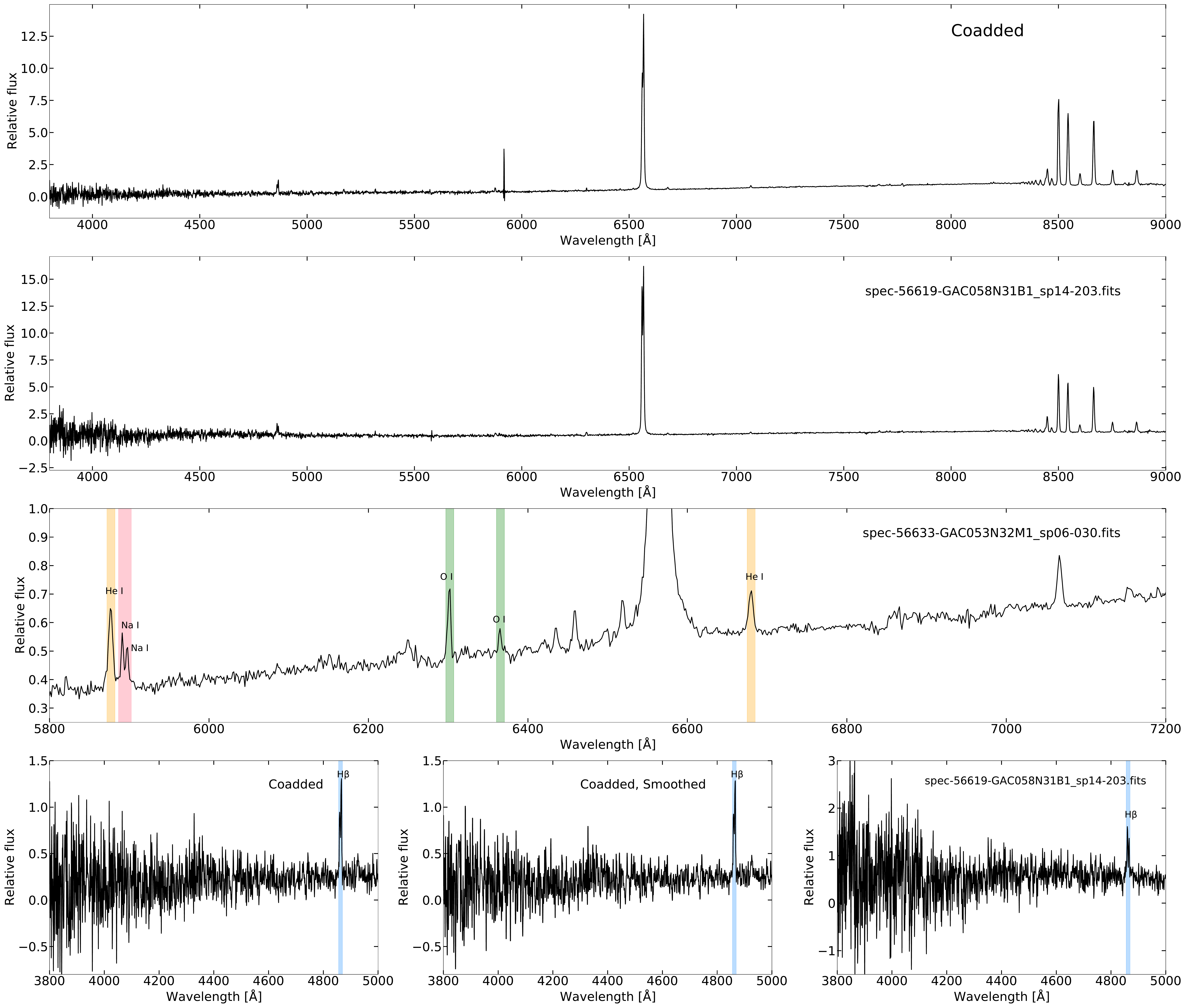}
\caption{The single-exposure and coadded spectra of \#21 J034344.48+314309.3, an outbursting EXor Herbig star, possible precursor of a Herbig Ae star.
\label{fig:fig8}}
\end{figure*}

Seven low-resolution spectra for \#21 are available from LAMOST DR8 observed between January 28 and December 6 in 2013. The single-exposure and coadded spectra are shown as Figure \ref{fig:fig8}. Besides the intense emission lines of $\rm H\alpha$, H\Rmnum{1} Paschen series, and Ca\Rmnum{2} triplet, we also detected obvious He\Rmnum{1} 5877, 6680, O\Rmnum{1} 6302.046, 6365.536, and Na\Rmnum{1} 5891.583, 5897.558. Typical evidences, especially the intense H\Rmnum{1}, He\Rmnum{1}, and Na\Rmnum{1} emission lines, clearly reveal the sudden increase of the mass accretion rate. When inspecting the blue arm spectra, the blurry photospheric continuum is hard to be discerned clearly no matter with the single-exposure one, the coadded one, and the coadded smoothed one. We suggest the flux at optical wavelength is dominated by its accretion disk during the outbursts. As pointed out by \citet{1985ApJ...299..462H}, rapid accretion onto PMS stars may cause the outburst, and emission from the hot, optically thick accretion disk dominates the system light at maxium. We found the direct evidence of outbursts using the LAMOST spectra, supporting the EXor nature proposed by \citet{1.2014ApJ...782...51A}.

Due to the intense emissions originated from the accretion disk, no credible stellar parameter can be determined for \#21 using the LAMOST spectra. Because the outbursts and EXor nature were taken into account, we prefer the stellar parameters ($\rm T_{eff}$ as 5718--5209 K, initial mass as 2.1--1.5 $M_\odot$, stellar luminosity as 13--5.4 $L_\odot$) determined by \citet{1.2014ApJ...782...51A}. Moreover, the determined $\rm T_{eff}$ of 5718 K is also consistent with the G-type nature at optical wavelengths during the outbursts suggested by \citet{1985ApJ...299..462H}. Besides, the G-type assumption is also reasonable according to our blurry photospheric continuum. Thus, we suggest \#21 is an outbursting EXor Herbig star, possible precursor of a Herbig Ae star.

\#21 is associated with the Perseus Molecular Cloud \citep{1.2013ApJS..205....5H,1.2009ApJS..181..321E}. According to the parallax provided by Gaia, the determined distance to \#21 is about 305.7 pc. As described in Section \ref{subsec:infrared}, the infrared spectrum reveals the existence of silicate, both amorphous and crystalline. Similarly, \citet{2009Natur.459..224A} detected crystalline forsterite features in the outburst spectrum of EX Lupi, the prototype EXors. They pointed out the crystalline features were not present in quiescence, and concluded that the crystals were produced through thermal annealing in the surface layer of the inner disk by heat from the outburst.
%In stark contrast to the intense $\rm H\alpha$ emission, the $\rm H\beta$ appears much gentler. Serious extinction accounts for the different qualities of the blue and red spectra. The evidence above reveals the existence of a warm dust circumstellar disk encompassing the hot gaseous inner disk. Using the same method in Section \ref{subsec:extinction}, we set the $\rm T_{eff}$ range as 10,000$-$30,000 K (B type) and estimated the parameters for \#21 as $A_V=12$, $T_{eff}=16,000 K$, $L=571.2 L_\odot$, $M=4.95 M_\odot$, and $Age=0.73 Myr$. We also calculated the spectral index $\alpha$, as described in detail in Section \ref{subsec:SED}. With the slope of $\alpha=-0.442$, \#21 is a Class \Rmnum{2} YSO near the boundary of the flat-spectrum class ($-0.3\le \alpha < 0.3$), which has good coherence with its Herbig Be nature with large extinction. The derived parameters for \#21 here are rough and they should be taken prudently, because its effective temperature cannot be confined into a more accurate range using the low-resolution spectra with the large extinction.

\subsection{Comparison with Herbig Ae/Be stars} \label{subsec:comparsion}

\subsubsection{The NIR and MIR spectral indices} \label{subsec:spectral indices}

According to the pre-main-sequence evolutionary tracks, most of the F-type PMS stars will eventually become HAeBes. To make a statistic analysis on the disk properties of Herbig stars, we also collected the known HAeBes and their young precursors with determined parameters. Besides our 20 F-type Herbig stars, 429 archival Herbig stars were also retrieved, of which 252 are from \citet{252HAeBe.2018A&A...620A.128V}, 128 are from \citet{2022ApJ...930...39V}, and 49 are from \citet{2021A&A...652A.133V}. We obtained infrared photometric data from the AllWISE. We removed sources with `dubious' marks or detected saturations, and finally got 365 Herbig stars with determined parameters for the following statistic analysis.
%To analyze the difference between the two samples, we also included the 72 HAeBes (71 confirmed and one credible candidate) from \citet{2022ApJS..259...38Z} based on LAMOST DR7 as a contrast.

\begin{figure*}
\includegraphics[width=\textwidth]{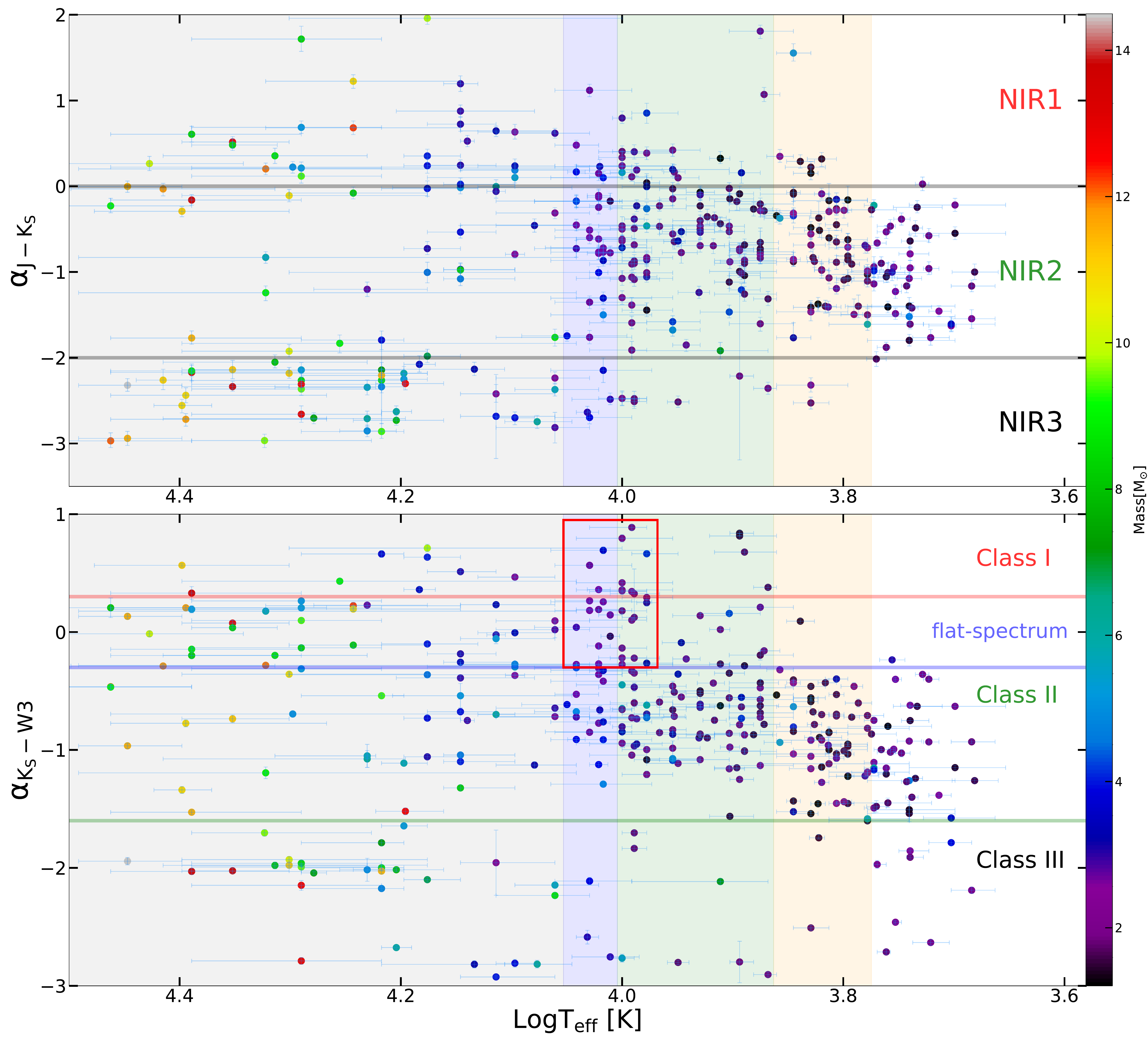}
\caption{The IR spectral indices of Herbig stars ($\alpha_{J-K_S}$ for the NIR and $\alpha_{K_S-W3}$ for the MIR). The Herbig stars and the determined parameters are from \citet{252HAeBe.2018A&A...620A.128V,2022ApJ...930...39V,2021A&A...652A.133V} and this work. The colorbar represents stellar masses, and we only presented the 354 stars with stellar masses less than 15 $M_\odot$ for better presentation of the lower side. The gray lines ($\alpha_{J-K_S}=0$, $\alpha_{J-K_S}=-2$) in the upper panel are the boundaries of NIR1, NIR2, and NIR3 classes. The light red, blue, and green lines ($\alpha_{K_S-W3}=0.3$, $\alpha_{K_S-W3}=-0.3$, $\alpha_{K_S-W3}=-1.6$, respectively) in the bottom panel are the boundaries of the modified 4-class YSO system defined by \citet{SED.1994ApJ...434..614G}. The background colors of `B8 and earlier', `B9 and B9.5', `A', `F', and `G and K' are set as light gray, blue, green, orange, and white, respectively.
\label{fig:fig9}}
\end{figure*}

Not only using the mid-infrared (MIR) spectral index $\alpha_{K_S-W3}$, we also calculated the near-infrared (NIR) spectral index $\alpha_{J-K_S}$ for each star. NIR emission arises from relatively large dust grains ($\ge1$ $\mu m$ for HAeBes) heated to the sublimation temperature 1500 K at the dust sublimation radius \citep{2002ApJ...579..694M}. MIR emission mainly arises from the warm dust. W3 band is most sensitive to 250 K blackbody dust, though contributions from dust with lower or higher temperature can also be present. The uncertainties of the spectral indices were derived from the photometric uncertainties. We plotted the spectral indices versus $\rm T_{eff}$ diagram as shown in Figure \ref{fig:fig9}. The colorbar represents stellar masses. In consideration of the color depth and visual perception, we only presented the 354 stars with stellar masses less than 15 $M_\odot$ for better presentation of the lower side. The light red, blue, and green lines ($\alpha_{K_S-W3}=0.3$, $\alpha_{K_S-W3}=-0.3$, $\alpha_{K_S-W3}=-1.6$, respectively) are the boundaries of the modified 4-class YSO system defined by \citet{SED.1994ApJ...434..614G} (Class \Rmnum{1}, flat-spectrum class, Class \Rmnum{2}, and Class \Rmnum{3}, Section \ref{subsec:SED} for details). The gray lines ($\alpha_{J-K_S}=0$, $\alpha_{J-K_S}=-2$) in the upper panel are the boundaries of NIR1, NIR2, and NIR3 classes. We divided the Herbig stars into five categories according to $\rm T_{eff}$. Statistics based on $\alpha_{K_S-W3}$ and $\alpha_{J-K_S}$ are shown as Table \ref{tab:counts}. Percentages in brackets represent the number ratios of Herbig stars within the same $\rm T_{eff}$ ranges. Considering the uncertainties of $\rm T_{eff}$, most of the `B9 and B9.5' type Herbig stars may be Herbig Ae stars. Thus, we set the `B9 and B9.5' type as a separate category in contrast to the `B8 and earlier' type. Intensive discussions on the five categories are presented in the following paragraphs.

%Moreover, as pointed by \citet{2014ApJ...797..112C,2015ApJ...810....5C}, Herbig Be stars (hereafter HBes, mid-to-early B-type stars) may accrete via a boundary layer rather than along the magnetic filed lines as that of Herbig Ae stars (hereafter HAes, A-type stars along with late B-type stars). Considering the possible different accretion mechanisms between HAes and HBes,

As seen in Figure \ref{fig:fig9} and Table \ref{tab:counts}, the spectral indices of F-type Herbig stars share the similar distribution with that of G- and K- type Herbig stars. The majority of these precursors (F-, G-, and K- type) have moderate NIR and MIR excesses, and belong to the Class \Rmnum{2} and NIR2 class. The relative large proportion of the Class \Rmnum{3} `G and K' type Herbig stars may be due to the small sample size. Besides, except for the proportion of `Class \Rmnum{1} \& flat-spectrum class', the distribution of `B9 and B9.5' type Herbig stars is much more similar to that of Herbig Ae stars than the distribution of `B8 and earlier' type.

There is a trend that the proportion of stars with moderate IR excesses (Class \Rmnum{2} and NIR2 class) decreases as $\rm T_{eff}$ increases. On the contrary, the proportions of stars with large (Class \Rmnum{1} \& flat-spectrum class and NIR1 class) or little (Class \Rmnum{3} and NIR3 class) IR excesses increase simultaneously with increasing $\rm T_{eff}$ in general. The trend becomes obvious when compare Herbig Be stars with Herbig Ae stars. Unlike Herbig Ae stars and the precursors, the majority of `B8 and earlier' type Herbig stars have large or little IR excesses instead of the moderate ones. In the 113 `B8 and earlier' type Herbig stars, 28.3\% and 42.5\% were classified as the Class \Rmnum{3} and NIR3 class separately (little IR excess), and 39.8\% and 30.1\% were classified as the `Class \Rmnum{1} \& flat-spectrum class' and NIR1 class separately (large IR excess). Moreover, 25.7\% (29/113) were classified as both Class \Rmnum{3} and NIR1 class, and 22.1\% (25/113) were classified as both `Class \Rmnum{1} \& flat-spectrum class' and NIR1 class, in the 113 `B8 and earlier' type Herbig stars. We suspect that the Herbig stars with little IR excesses may likely just reach the main sequence with dust disk dispersed by stellar radiation. The great proportions (28.3\% for the Class \Rmnum{3} and 42.5\% for the NIR3 class) are also consistent with the fact that more massive stars evolve much faster and approach the main-sequence phase more quickly.

% 283 Herbig stars have $\rm T_{eff}$ below 12,300 K (hereafter called HAes and the precursors), of which 71.7\% (203/283) belong to the Class \Rmnum{2}, 18.0\% (51/283) belong to the Class \Rmnum{1} or the flat-spectrum class, and 10.3\% (29/283) belong to the Class \Rmnum{3}, according to the MIR spectral index $\alpha_{K_S-W3}$. Besides, 79.9\% (226/283) of the HAes and the precursors have the NIR spectral index $\alpha_{J-K_S}$ ranging from -2 to 0. In general, most of the HAes and the precursors have moderate MIR and NIR excess.

Hotter Herbig stars tend to have a larger proportion with large IR excesses described as above. It may be due to the fact that hotter stars have larger areas of re-emitting dust, although there is some scatter due to the particularities of each disk. Because of the high stellar luminosity and the intense stellar radiation from Herbig Be stars, distant warm dust may also contribute to the MIR excess. Similarly, the high stellar luminosity of Herbig Be stars also account for the large dust sublimation radius along with the large NIR excess. As revealed by \citet{2002ApJ...579..694M,2008A&A...489.1157K}, the NIR emission likely traces dust at the dust sublimation radius in Herbig stars, and the measured NIR sizes are closely related to the stellar luminosity. The high $\rm T_{eff}$ and stellar luminosity usually lead to the large radius of optical thin cavity surrounding the star. \citet{2009A&A...502L..17A} also pointed out that an inflation of the inner disk height naturally leads to a higher NIR excess.

The proportions of precursors with large IR (especially MIR) excesses are much smaller than that of Herbig Ae stars. The only 2 precursors with $\alpha_{K_S-W3}$ above -0.3 were both classified as the flat-spectrum class, without any Class \Rmnum{1} star. In addition to the cause mentioned above (areas of re-emitting dust), an observational effect may also contribute to it. Those most embedded precursors with large IR excesses and low $\rm T_{eff}$ are hard to be detected in general.

%The only difference between the distributions of HAes and the precursors is the percentage of stars with large MIR excess. 49 of the 51 HAes and the precursors with $\alpha_{K_S-W3}$ above -0.3 (Class \Rmnum{1} or the flat-spectrum class) are with $\rm T_{eff}$ above 7300 K (earlier than F0).

%The number ratio of HBes with large NIR excess to HBes with moderate NIR excess (1.14, 33/29) is much larger than that of HAes and the precursors (0.16, 36/226). Besides, the spacial distribution of dust grains plays an important role. We suggest the complicated environment and the relative small sample of HBes may be the main reasons for the random distribution of $\alpha_{K_S-W3}$.

In this work, \#13 is the only one with $\alpha_{K_S-W3}$ above -0.3 and its determined $\rm T_{eff}$ is 6900 K. Despite taking a large value of $R_V$ ($R_V=5$) as suggested by \citet{2004AJ....127.1682H} for \#13 rather than the standard $R_V=3.1$, we might still underestimate its extinction and luminosity if the real $R_V$ was even larger. Similar to \#13, 29 of 32 Herbig stars with $\rm T_{eff}$ around 10,000 K (9500$-$11,300 K), marked in the red square in Figure \ref{fig:fig9}, have the determined stellar masses less than 3 $M_\odot$ and were classified as the Class \Rmnum{1} or the flat-spectrum class. Many of them have the determined ages older than the very young nature indicated by the early stage of disk evolution suggested by \citet{SED.1994ApJ...434..614G,1994ApJ...434..330C}. It should be noted that disk evolutionary theories are varied and nowadays often present several paths for disk evolution. However, according to the pre-main-sequence evolutionary tracks, the 29 stars are close to the main sequence and should not have, in principle, IR excesses that large. As pointed out by \citet{1989ApJ...345..245C}, the derived mean extinction law depends on only one parameter, which is chosen to be $R_V \equiv A_V/E(B-V)$. The value of $R_V$ is relevant to the sizes and distributions of dust grains. $R_V=3.1$ is the standard value for the diffuse interstellar medium and $R_V=5$ is a value found in some dense clouds. \citet{2004ApJ...609..589W} also pointed out preferential removal of small dust grains such as by coagulation in dense molecular clouds will result in a gray extinction law with large $R_V$. \citet{2004AJ....127.1682H} suggested the value of $R_V$ can be used to infer grain properties of the dust surrounding HAeBe stars. We suggest the average Galactic value $R_V=3.1$ is credible in the less embedded Herbig stars but not the embedded ones. The Herbig stars with extremely large IR excesses (Class \Rmnum{1} \& flat-spectrum class) may be embedded in dense molecular clouds where significant grain growth happens and large size dust grains dominate. We suggest the assumption of $R_V=3.1$ for the embedded Herbig stars with extremely large IR excesses during the parameter determination in \citet{252HAeBe.2018A&A...620A.128V,2022ApJ...930...39V} may result in the underestimation of the extinction, stellar luminosity, and mass. Large values of $R_V$ ($R_V=5$ or larger) should be taken in the cases of `Class \Rmnum{1} \& flat-spectrum class' stars with large IR excesses, and these stars should have larger extinctions and stellar luminosities. According to the pre-main-sequence evolutionary tracks, they should be much younger and more massive than previously thought.

Our simplified indices can trace the general IR trends of Herbig stars rather than the refined disk structures. In fact, the detected disk structures of Herbig stars are complicated, such as cavities, gaps, and spiral arms \citep{2019ApJ...871...48P,1.spiral.2018AJ....156...63U,1.spiral.2016AJ....152..222A}. \citet{2001A&A...365..476M} defined two groups (Group \Rmnum{1} and Group \Rmnum{2}) for HAeBes based on IR excess. The continuum of Group \Rmnum{1} can be reconstructed by a power-law and a black body, and that of Group \Rmnum{2} only needs a power-law to fit. MIR excess is dominant and rising for Group \Rmnum{1}, while moderate and rather descending for Group \Rmnum{2}. \citet{2001A&A...365..476M,group.2004A&A...417..159D,group.2004A&A...421.1075D,group.2005A&A...434..971D} suggested Group \Rmnum{1} may evolve into Group \Rmnum{2} through grain growth and/or settling. However, \citet{2015A&A...581A.107M,2017A&A...603A..21G} argued Group \Rmnum{2} may involve into Group \Rmnum{1} through formation of (giant) planets. Recently, \citet{2021A&A...652A.133V} pointed out Group \Rmnum{1} and Group \Rmnum{2} are disconnected and represent two different evolutionary paths. Moreover, a number of recent work analysed the protoplanetary disks of Herbig stars based on the data from the Atacama Large Millimeter/submillimeter Array (ALMA), the Very Large Telescope Interferometer (VLTI) GRAVITY, and the Gemini- Large Imaging with GPI Herbig/T-tauri Survey (Gemini-LIGHTS) \citep{new.2022arXiv220605815R,new.2022A&A...658A.183B,new.2022A&A...658A.112S,new.2020A&A...642A.164V,new.2019A&A...626A..11C}. Results reveal that the dust spatial distributions of Herbig stars are even more complicated and show large star-to-star differences. We simplified the situation and made the statistic analysis of all known Herbig stars with NIR and MIR spectral indices in this section. Considering the complicated situation of Herbig stars, following detailed investigations for individuals are necessary to verify the trends and our speculations mentioned above.

{
\setlength\LTleft{-1in}
\setlength\LTright{-1in plus 1 fill}
\setlength{\tabcolsep}{5pt}

\small
\tiny
\begin{deluxetable*}{ccccccccc}
\tablenum{3}
\tablecaption{Classification of Herbig stars based on $\alpha_{K_S-W3}$ and $\alpha_{J-K_S}$ \label{tab:counts}}
\tablewidth{0pt}
\tablehead{
\colhead{ } & \colhead{$\rm T_{eff}$} & \colhead{ } & \colhead{Class \Rmnum{1} \& } & \colhead{ } & \colhead{ } & \colhead{ } & \colhead{ } & \colhead{ }\\
\colhead{Spectral type} & \colhead{[K]} & \colhead{Count} & \colhead{flat-spectrum class} & \colhead{Class \Rmnum{2}} & \colhead{Class \Rmnum{3}} & \colhead{NIR1} & \colhead{NIR2} & \colhead{NIR3}
}
\startdata
 B8 and earlier & $>$ 11,300 & 113 & 45 (39.8\%) & 36 (31.9\%) & 32 (28.3\%) & 34 (30.1\%) & 31 (27.4\%) & 48 (42.5\%) \\
 B9 and B9.5 & 10,100$-$11,300 & 34 & 14 (41.2\%) & 17 (50.0\%) & 3 (8.8\%) & 6 (17.6\%) & 24 (70.6\%) & 4 (11.8\%) \\
 A & 7300$-$10,100 & 112 & 32 (28.6\%) & 73 (65.2\%) & 7 (6.2\%) & 22 (19.6\%) & 83 (74.1\%) & 7 (6.3\%) \\
 F & 6000$-$7300 & 61 & 1 (1.6\%) & 57 (93.5\%) & 3 (4.9\%) & 6 (9.8\%) & 53 (86.9\%) & 2 (3.3\%) \\
 G and K & $<$6000 & 45 & 1 (2.2\%) & 35 (77.8\%) & 9 (20.0\%) & 1 (2.2\%) & 42 (93.4\%) & 2 (4.4\%)
\enddata
\tablecomments{The `Count' column is the total number of Herbig stars with corresponding spectral types and $\rm T_{eff}$. Columns 4$-$9 show the counts of Herbig stars with specified classes. Percentages in brackets are the number ratios of Herbig stars within the same $\rm T_{eff}$ ranges. The 4-class system (Class \Rmnum{1}, flat-spectrum class, Class \Rmnum{2}, Class \Rmnum{3}) is based on $\alpha_{K_S-W3}$, and the 3-class one (NIR1, NIR2, NIR3) is based on $\alpha_{J-K_S}$.
}
\end{deluxetable*}
}

\subsubsection{The detection frequency of PAH} \label{subsec:PAH}

Statistic studies reveal a correlation between the detection frequency of PAH and $\rm T_{eff}$. \citet{PAH.2010ApJ...718..558A} found 37 of the 53 Herbig Ae stars (70\%) display PAH emission. \citet{HAeBe.2019ApJ...878..147K} pointed out all the 13 HAeBes identified in the Small Magellanic Cloud show PAH emission. \citet{PAH.2006A&A...459..545G} detected PAH features in 3 of the 38 T Tauri stars (8\%, with an additional 14 tentative detections). \citet{2021A&A...652A.133V} pointed out 27\% of the 49 IMTT stars show PAH features, and the detection rate becomes 44\% when considering tentative detections. In this work, only 4 of the 20 F-type Herbig stars have Spitzer IRS spectra, three of which with $\rm T_{eff}$ about 6400 K display extremely weak PAH emission. As is suggested by \citet{2021A&A...652A.133V}, the detection frequency tends to decrease when the $\rm T_{eff}$ and the UV flux decrease. Despite our small sample, no intense PAH emission detected also supports their suggestion.

\section{Summary} \label{sec:summary}

In this work, we used the 2.53 million F-type spectra from LAMOST DR8 and identified 20 F-type Herbig stars along with 22 pre-main-sequence candidates. The effective temperature, distance, extinction, stellar luminosity, mass, and radius were derived for each Herbig star based on optical spectra, photometry, Gaia EDR3 parallaxes, and pre-main-sequence evolutionary tracks.

According to SEDs, 19 F-type Herbig stars belong to Class \Rmnum {2} YSOs, and one belongs to the flat-spectrum class. Four have Spitzer IRS spectra, of which three show extremely weak PAH emissions, and three with both amorphous and crystalline silicate emissions share the similar parameters and are at the same evolutionary stage.

We detected an additional Herbig star J034344.48+314309.3 (R.A. 55.935352, Decl. 31.719259). It was thought to be an EXor, a T Tauri star or a Be star. It is a solar-nearby star with a distance of 305.7 pc. We detected intense emission lines of H\Rmnum{1}, He\Rmnum{1}, O\Rmnum{1}, Na\Rmnum{1}, and Ca\Rmnum{2} in its optical spectra, originated from the rapid accretion during the outbursts. We also detected silicate emission features (both amorphous and crystalline) in its infrared spectrum. We suggest it is an outbursting EXor Herbig star, possible precursor of a Herbig Ae star.

We also made a statistic analysis on the disk properties of all known Herbig stars using the defined infrared spectral indices ($\alpha_{J-K_S}$ and $\alpha_{K_S-W3}$). The possible trends and our speculations are listed as bellow.
\begin{enumerate}
\item The spectral indices of F-type Herbig stars share the similar distribution with that of G- and K- type Herbig stars. The majority of these precursors (F-, G-, or K- type) have moderate NIR and MIR excesses, and belong to the Class \Rmnum{2} and NIR2 class. The distribution of `B9 and B9.5' type Herbig stars is much more similar to that of Herbig Ae stars than the distribution of `B8 and earlier' type.
\item The proportion of Herbig stars with moderate IR excesses decreases as $\rm T_{eff}$ increases. We suspect that the Herbig stars with little IR excesses may likely just reach the main sequence with dust disk dispersed by stellar radiation.
\item Hotter Herbig stars tend to have a larger proportion with large IR excesses. It may be due to the fact that hotter stars have larger areas of re-emitting dust, although there is some scatter due to the particularities of each disk.
\item The proportions of precursors with large IR (especially MIR) excesses are much smaller than that of Herbig Ae stars. In addition to the cause mentioned above (areas of re-emitting dust), an observational effect may also contribute to it. Those most embedded precursors with large IR excesses and low $\rm T_{eff}$ are hard to be detected in general.
\item We suggest the average Galactic value $R_V=3.1$ is credible in the less embedded Herbig stars but not the embedded ones. Large values of $R_V$ ($R_V=5$ or larger) should be taken in the cases of `Class \Rmnum{1} \& flat-spectrum class' stars with large IR excesses, and these stars should have larger extinctions and stellar luminosities. According to the pre-main-sequence evolutionary tracks, they should be much younger and more massive than previously thought.
\end{enumerate}

\section*{Acknowledgements}

We thank the anonymous reviewer and Aigen Li for suggestions and comments that significantly improved the paper. This work is supported by National Science Foundation of China (Nos U1931209, 12003050, 12133002) and National Key R\&D Program of China (No. 2019YFA0405502). Guoshoujing Telescope (the Large Sky Area Multi-Object Fiber Spectroscopic Telescope, LAMOST) is a National Major Scientific Project built by the Chinese Academy of Sciences. Funding for the project has been provided by the National Development and Reform Commission. LAMOST is operated and managed by the National Astronomical Observatories, Chinese Academy of Sciences. {This publication makes use of data products from the Wide-field Infrared Survey Explorer, which is a joint project of the University of California, Los Angeles, and the Jet Propulsion Laboratory/California Institute of Technology, funded by the National Aeronautics and Space Administration. This research makes use of data from the European Space Agency (ESA) mission Gaia, processed by the Gaia Data Processing and Analysis Consortium. This publication makes use of VOSA, developed under the Spanish Virtual Observatory project funded by MCIN/AEI/10.13039/501100011033/ through grant PID2020-112949GB-I00. VOSA has been partially updated by using funding from the European Union's Horizon 2020 Research and Innovation Programme, under Grant Agreement n$^\circ$ 776403 (EXOPLANETS-A). This research also makes use of Astropy, a community-developed core Python package for Astronomy \citep{astropy2013A&A...558A..33A}, the TOPCAT tool \citep{TOPCAT2005ASPC..347...29T} and the VizieR catalog access tool and the Simbad database, operated at CDS, Strasbourg, France.}

\appendix

%\begin{appendices}
\section{Pre-main-sequence candidates}

Data for the 22 Pre-main-sequence candidates is provided in Table \ref{tab:candidates}.

{
\setlength\LTleft{-1in}
\setlength\LTright{-1in plus 1 fill}
\setlength{\tabcolsep}{5pt}

\small
\tiny
\begin{deluxetable*}{ccccc}
\tablenum{A}
\tablecaption{Pre-main-sequence candidates\label{tab:candidates}}
\tablewidth{0pt}
\tablehead{
\colhead{No.} & \colhead{Designation} & \colhead{R.A. [degree]} & \colhead{Decl. [degree]} & \colhead{Specname}
}
\startdata
 C1 & J032025.39+554621.1 & 50.105798 & 55.772531 & spec-56937-HD032616N560830V01\_sp03-031.fits \\
 C2 & J034512.81+521437.7 & 56.303389 & 52.243824 & spec-55903-B90304\_sp11-014.fits \\
 C3 & J034714.72+472557.6 & 56.811339 & 47.432686 & spec-56627-GAC056N46V1\_sp09-088.fits \\
 C4 & J044319.47+445421.7 & 70.831162 & 44.906046 & spec-57425-GAC067N44B3\_sp13-045.fits \\
 C5 & J051905.92+383946.8 & 79.774677 & 38.663002 & spec-57015-GAC082N38M1\_sp14-233.fits \\
 C6 & J053304.15+384300.9 & 83.267325 & 38.716921 & spec-55976-GAC\_084N40\_V1\_sp05-223.fits \\
 C7 & J053715.40+244917.2 & 84.314167 & 24.821472 & spec-56271-GAC084N26B1\_sp01-054.fits \\
 C8 & J053813.35+274047.1 & 84.555634 & 27.67975 & spec-55918-GAC\_082N29\_M1\_sp07-094.fits \\
 C9 & J054022.60+355913.0 & 85.094182 & 35.986965 & spec-57328-GAC084N35B1\_sp09-202.fits \\
 C10& J054950.53+270701.4 & 87.460583 & 27.117083 & spec-55876-GAC\_089N28\_B1\_sp02-147.fits \\
 C11& J055018.51+281042.6 & 87.577164 & 28.178522 & spec-55876-GAC\_089N28\_B1\_sp10-167.fits \\
 C12& J055532.35+274451.3 & 88.884792 & 27.747611 & spec-55892-GAC\_082N27\_M1\_sp09-147.fits \\
 C13& J055801.69+192238.7 & 89.507083 & 19.377417 & spec-56648-GAC088N20B1\_sp08-116.fits \\
 C14& J055828.98+264039.9 & 89.620756 & 26.677767 & spec-55892-GAC\_082N27\_M1\_sp06-207.fits \\
 C15& J060413.19+211639.6 & 91.054997 & 21.277676 & spec-56595-GAC088N22M1\_sp07-143.fits \\
 C16& J061136.87+203104.0 & 92.903638 & 20.517795 & spec-57753-GAC093N19M2\_sp15-198.fits \\
 C17& J061141.85+203158.9 & 92.924376 & 20.533032 & spec-57753-GAC093N19M2\_sp15-193.fits \\
 C18& J061624.46+241837.1 & 94.101938 & 24.310332 & spec-57009-GAC091N23M1\_sp13-002.fits \\
 C19& J063144.07+021335.0 & 97.933625 & 2.2263889 & spec-58137-GAC099N02B2\_sp10-211.fits \\
 C20& J063146.08+044228.8 & 97.942032 & 4.7080234 & spec-55976-GAC\_099N04\_V5\_sp10-243.fits \\
 C21& J063856.02+542940.4 & 99.733451 & 54.49457 & spec-57814-GAC101N53M1\_sp16-164.fits \\
 C22& J212547.34-022251.2 & 321.447259& -2.380904 & spec-57641-EG213415S032438V01\_sp14-106.fits
\enddata
\tablecomments{The `No.' column is the sequential identifier of the 22 Pre-main-sequence candidates. The `Specname' column is the FITS name of each spectrum.
}
\end{deluxetable*}
}

\bibliography{ref}{}
\bibliographystyle{aasjournal}



\end{document}